\documentclass{emulateapj}
\usepackage{natbib,graphicx,url}
\slugcomment{{\sc Accepted for Publication in the Astronomical Journal:}}
\shorttitle{SDSS BL Lac Candidates}
\shortauthors{Plotkin et al.}

\begin{document}

\def\bl{BL~Lac}
\defcitealias{collinge05}{C05}

\makeatletter
\newcommand\Rom[1]{\@Roman{#1}}  
\newcommand\rom[1]{\@roman{#1}}  
\makeatother

\newcommand{\noptfeat}{1310}    
\newcommand{\noptall}{723}         
\newcommand{\noptrl}{637}           
\newcommand{\noptrlHigh}{613}   
\newcommand{\noptrlLow}{24}      
\newcommand{\noptrq}{86}           
\newcommand{\noptrqHigh}{56}    
\newcommand{\noptrqLow}{30}     

\newcommand{\noptallXray}{294}   
\newcommand{\noptallNoXray}{429}  
\newcommand{\noptallXrayNoRadio}{9} 
\newcommand{\noptallFirst}{561}    
\newcommand{\noptallNvss}{554}   
\newcommand{\noptallRadio}{599}  
\newcommand{\noptallFirstNOTINNvss}{45}   
\newcommand{\noptallNvssNOTINFirst}{38}   
\newcommand{\noptallFirstANDNvss}{516}  
\newcommand{\noptallRadioLimits}{117}     
\newcommand{\noptallNoRadioInfo}{7}    
\newcommand{\noptrlXray}{289}   
\newcommand{\noptrlXrayAndRadio}{284}  
\newcommand{\noptrlNoXray}{348}  
\newcommand{\noptrlXrayNoRadio}{5} 
\newcommand{\noptrlFirst}{552}    
\newcommand{\noptrlRadio}{590}  
\newcommand{\noptrlNvssNOTINFirst}{38}   
\newcommand{\noptrlRadioLimits}{40}     
\newcommand{\noptrlNoRadioInfo}{7}    

\newcommand{\noptallNoz}{275}       
\newcommand{\noptallZok}{448}
\newcommand{\noptallZrel}{269}
\newcommand{\noptallZtent}{130}
\newcommand{\noptallZll}{49}
\newcommand{\noptallZtentPlusZrel}{399}
\newcommand{\noptrlNoz}{270}       
\newcommand{\noptrlZok}{367}
\newcommand{\noptrlZrel}{231}
\newcommand{\noptrlZtent}{94}
\newcommand{\noptrlZll}{42}
\newcommand{\noptrqNoz}{5}       
\newcommand{\noptrqZok}{81}
\newcommand{\noptrqZrel}{38}
\newcommand{\noptrqZtent}{36}
\newcommand{\noptrqZll}{7}
\newcommand{\noptRquietLowZ}{13}       
\newcommand{\noptRquietMidZ}{55}       
\newcommand{\noptRquietBigZ}{13}       
\newcommand{\noptrqWLQ}{5}                

\newcommand{\noptallHGpref}{314}    
\newcommand{\noptallPLpref}{409}     

\newcommand{\noptRquietXray}{5}       
\newcommand{\noptRquietRadioDetection}{9}   
\newcommand{\noptRquietRadioLimit}{77}    

\newcommand{\noptrej}{587}        
\newcommand{\noptrejdc}{500}   
\newcommand{\noptrejbgpm}{4}   
\newcommand{\noptrejlrg}{83}     

\newcommand{\tabnwlq}{20}   
\newcommand{\tabnwlqBL}{6}  
\newcommand{\tabnwlqNotBL}{14}  

\title{Optically Selected BL Lacertae Candidates from the Sloan Digital Sky Survey Data Release Seven}

\author{
Richard~M.~Plotkin\altaffilmark{1},
Scott~F.~Anderson\altaffilmark{1},
W.~N.~Brandt\altaffilmark{2},
Aleksandar~M.~Diamond-Stanic\altaffilmark{3},
Xiaohui~Fan\altaffilmark{3},
Patrick~B.~Hall\altaffilmark{4},
Amy~E.~Kimball\altaffilmark{1},
Michael~W.~Richmond\altaffilmark{5},
Donald~P.~Schneider\altaffilmark{2},
Ohad~Shemmer\altaffilmark{6},
Wolfgang~Voges\altaffilmark{7,8},
Donald~G.~York\altaffilmark{9},
Neta~A.~Bahcall\altaffilmark{10},
Stephanie~Snedden\altaffilmark{11},
Dmitry~Bizyaev\altaffilmark{11},
Howard~Brewington\altaffilmark{11},
Viktor~Malanushenko\altaffilmark{11},
Elena~Malanushenko\altaffilmark{11},
Dan~Oravetz\altaffilmark{11},
Kaike~Pan\altaffilmark{11},
Audrey~Simmonds\altaffilmark{11}
}

\altaffiltext{1}{Department of Astronomy, University of Washington, Box 351580, Seattle, WA 98195, USA; plotkin@astro.washington.edu, anderson@astro.washington.edu}
\altaffiltext{2}{Department of Astronomy and Astrophysics, Pennsylvania Sate University, 525 Davey Laboratory, University Park, PA 16802, USA}
\altaffiltext{3}{Steward Observatory, University of Arizona, Tucson, AZ 85721, USA}
\altaffiltext{4}{Department of Physics and Astronomy, York University, 4700 Keele Street, Toronto, ON M3J 1P3, Canada}
\altaffiltext{5}{Physics Department, Rochester Institute of Technology, Rochester, NY 14623, USA}
\altaffiltext{6}{Department of Physics, University of North Texas, Denton, TX 76203, USA}
\altaffiltext{7}{Max-Planck-Institut f\"ur extraterrestrische Physik, Giessenbachstrasse, Postfach 1312, D-85741, Garching, Germany}
\altaffiltext{8}{Max Planck Digital Libray, Amalienstrasse 33, D-80799 M\"unchen, Germany}
\altaffiltext{9}{The University of Chicago and the Fermi Institute, 5640 South Ellis Avenue, Chicago, IL 60637, USA}
\altaffiltext{10}{Princeton University Observatory, Peyton Hall, Ivy Lane, Princeton, NJ 08544, USA}
\altaffiltext{11}{Apache Point Observatory, Sunspot, NM 88349, USA}

\begin{abstract}
We present a sample of \noptall\ optically selected \bl\ candidates from the SDSS DR7 spectroscopic database encompassing  8250~deg$^2$ of sky; our sample constitutes one of the largest uniform BL Lac samples yet derived. Each \bl\ candidate has a high-quality SDSS spectrum from which we determine spectroscopic redshifts for $\sim$60\% of the objects. Redshift lower limits are estimated for the remaining objects utilizing the lack of host galaxy flux contamination in their optical spectra; we find that objects lacking spectroscopic redshifts are likely at systematically higher redshifts.  Approximately 80\% of our \bl\ candidates match to a radio source in FIRST/NVSS, and $\sim$40\% match to a ROSAT X-ray source.  The homogeneous multiwavelength coverage allows subdivision of the sample into \noptrl\ radio-loud \bl\ candidates and \noptrq\ weak-featured radio-quiet objects. The radio-loud objects broadly support the standard paradigm unifying \bl\ objects with beamed radio galaxies. We propose that the majority of the radio-quiet objects may be lower-redshift ($z<2.2$) analogs to high-redshift weak line quasars (i.e., AGN with unusually anemic broad emission line regions).  These would constitute the largest sample of such objects, being of similar size and complementary in redshift to the samples of high-redshift weak line quasars previously discovered by the SDSS.  However, some fraction of the weak-featured radio-quiet objects may instead populate a rare and extreme radio-weak tail of the much larger radio-loud \bl\ population.  Serendipitous discoveries of unusual white dwarfs, high-redshift weak line quasars, and broad absorption line quasars with extreme continuum dropoffs blueward of rest-frame 2800~\AA\ are also briefly described.
\end{abstract}


\section{Introduction}
BL~Lacertae objects compose an especially rare subclass of Active Galactic Nuclei (AGNs) traditionally interpreted as  low-luminosity radio galaxies with a relativistic jet pointed toward the observer \citep[e.g., see][]{blandford78, urry95}.  Emission from the relativistically boosted jet dominates the observed flux, yielding featureless (or nearly featureless) optical spectra, strong radio and X-ray emission, strong polarization, flat radio spectra, and rapid variability across the entire electromagnetic spectrum \citep[e.g., see][]{kollgaard94,perlman01}.  Given their odd spectral characteristics, and the need to obtain optical spectra to confirm a candidate as a \bl\ object, the most efficient \bl\ selection algorithms tend to search for optical counterparts to strong radio and/or X-ray sources \citep[e.g., see][]{stickel91,stocke91,perlman98,laurent99,landt01,anderson07,padovani07,piranomonte07,turriziani07,plotkin08}.

Optically selected \bl\ samples have historically been extraordinarily difficult to assemble, but they are highly desired.  Systematic searches based only on optical characteristics would allow the exploration of \bl\ properties with minimal concern for  biases introduced by radio and X-ray surveys (which generally tend to have shallower flux sensitivities than optical surveys.)  Also, optically selected samples might reveal new populations of \bl\ objects, and they might even lead to serendipitous discoveries of other, perhaps new, classes of featureless objects. Unfortunately, the odd spectral characteristics of \bl\ objects conspire to make selection based solely on optical colors ineffective.  For example, only 4 \bl\ objects were identified in the Palomar-Green Survey for ultraviolet-excess objects \citep[PG,][]{green86}; another 3 PG sources initially misclassified as DC (i.e., featureless) white dwarfs were identified as \bl\ objects only after the {\it ROSAT} All Sky Survey \citep[RASS,][]{voges99,voges00} came online \citep{fleming93}.  

Attempts have been made to exploit other distinctive \bl\ properties, like polarization and flux variability, to derive optically selected samples.   Such techniques, however, require observing large areas of the sky at multiple epochs just to assemble manageable lists of candidates suitable for spectroscopic follow-up.  For example, it would seem that the strong polarization of \bl\ objects ($P\gtrsim2\%$) should allow for efficient identification of \bl\ candidates.  However, most attempts along these lines have so far been unsuccessful \citep{impey82_optsel,borra84,jannuzi93}, probably because the polarized flux is variable with a low duty cycle: for example, X-ray selected \bl\ objects from the {\it Einstein} Observatory Extended
Medium-Sensitivity Survey \citep[EMSS,][]{stocke91}  spend over half their time with $P<4\%$ \citep[][]{jannuzi94}.   

Not until we entered the age of large-scale spectroscopic digital sky surveys has optical selection of \bl\ objects been a realistic endeavor.  Traditionally, the primary limitation of efficient \bl\ recovery algorithms is the large number of false positives returned, even when radio and/or X-ray information is consulted.  Optical spectroscopy is then required to make firm \bl\ identifications.  However, with a digital survey such as the Sloan Digital Sky Survey \citep[SDSS,][]{york00}, this task is less daunting because the requisite spectra already exist.  Furthermore, the most relevant spectra can be searched within databases, allowing for automated removal of many of the contaminants that previously had to be culled manually (after a significant investment in telescope time.)  

The first successful attempt at assembling an optically selected sample is the 2QZ \bl\ survey \citep{londish02,londish07}, derived from  the 2-degree field (2dF) and the 6-degree field (6dF) quasi-stellar object Redshift Surveys \citep{croom04}.  They searched over $\sim$10$^3$~deg$^2$ and recovered 7 confident \bl\ objects.  All 7 objects were additionally identified with radio and X-ray sources (post-selection), and they also showed optical flux variations in a photometric monitoring campaign over 2002--2004 \citep{nesci05}.   An additional object ($z=0.494$) is identified as a radio-quiet/weak \bl\ candidate \citep{londish04}.  This source lacks both a radio and an X-ray detection (with a radio to optical flux ratio limit placing it firmly in the radio-quiet regime\footnote{Radio-quiet quasars are commonly defined as having radio to optical flux ratios $R<10$ \citep[e.g., see][]{kellermann89, stocke92}.   Throughout this paper, we adopt the approximately similar $\alpha_{ro}<0.2$, where $\alpha_{ro}$ is the radio-to-optical broad-band spectral index: $\alpha_{ro}=-\log(L_o/L_r)/5.08$, where $L_o$ and $L_r$ are optical and radio specific luminosities at rest-frames 5000~\AA\ and 5~GHz, respectively \citep{stocke85}.}),  
but its optical properties are similar to (radio-loud) \bl\ objects.  The recovery of only a single extragalactic object with a featureless spectrum lacking radio emission by the  2QZ survey confirms the \citet{stocke90} result that radio-quiet \bl\ objects must be extremely rare if they even exist at all \citep[also see][]{jannuzi93}.  

Searching 2860~deg$^2$, \citet{collinge05}, hereafter \citetalias{collinge05}, assembled a larger optically selected sample from the SDSS.  Their sample contains 386 nearly featureless spectra, including 240 {\it probable} \bl\ candidates and 146 {\it possible} \bl\ candidates (the latter are most likely DC white dwarfs, proper motions are consulted for their classification.)  The vast majority of the {\it probable} candidates either have firm extragalactic redshifts, or they match to a radio source in the Faint Images of the Radio Sky at Twenty-cm survey \citep[FIRST,][]{becker95} and/or the NRAO VLA Sky Survey \citep[NVSS,][]{condon98},  or to a RASS X-ray source (correlations to multiwavelength catalogs are performed post-selection.)  

\citetalias{collinge05} also present a list of 27 intriguing radio-quiet/weak \bl\ candidates.   These objects have nearly featureless optical spectra, and their radio fluxes (or limits in the cases of radio non-detections) place their radio to optical flux ratios within or tantalizingly close to the radio-quiet regime.   All but one of these also lack X-ray detections in RASS (but their extant X-ray limits are not sensitive enough to declare them X-ray weak.)  Some of these objects are at high-redshift ($z>2.2$), and those high-redshift objects may be alternately described as members of a population of high-redshift weak line quasars (WLQs) discovered by the SDSS \citep[e.g., see][]{fan99,anderson01,shemmer06,shemmer09,diamond09}.  Three of these high-redshift objects were detected in follow-on {\it Chandra} X-ray observations by \citet{shemmer09}, and they are X-ray weaker than typical radio-loud \bl\ objects.   Other radio-quiet \bl\ candidates in \citetalias{collinge05} have lower redshifts, and it is unclear if they are more closely related to \bl\ objects or to WLQs.  Regardless of their true nature, \citetalias{collinge05} demonstrate that optical selection of \bl\ objects from the SDSS is efficient,  producing a large enough sample to reveal especially rare sub-populations of objects with featureless spectra in numbers comparable to entire venerable radio and X-ray selected \bl\ samples.  

Here, we expand on \citetalias{collinge05} and continue the search for \bl\ objects in the SDSS (which now covers almost three times the sky area as the \citetalias{collinge05} sample).  We recover \noptall\ objects.  Our automated selection algorithm is described in \S \ref{sec:ch4_sampsel}.  We visually inspect $\sim$23,000 spectra, and we discuss the removal of various classes of contaminants (including stars, post-starburst galaxies, quasars, etc.)\ in \S \ref{sec:ch4_visinspec}; we also present lists of serendipitously discovered unusual white dwarfs, WLQs, and unusual broad absorption line quasars.  In \S \ref{sec:ch4_postvisinspec} we describe the removal of contaminants that survive visual inspection (including featureless white dwarfs and large red galaxies.)  The final sample is presented in \S \ref{sec:ch4_finalsamp}, and we describe a spectral decomposition of the \bl\ host galaxy from the AGN.  We use this decomposition to measure optical spectral indices for the AGN, to estimate lower redshift limits by assuming \bl\ host galaxies are standard candles, and to calculate decomposed AGN component optical fluxes and luminosities.  We also correlate the \noptall\ \bl\ candidates to the FIRST and NVSS radio surveys, and to the RASS X-ray survey.   The sample is discussed in \S \ref{sec:ch4_discussion}, which focuses on a particularly interesting subset of \noptrq\ radio-quiet objects with weak spectral features.   Finally, our main conclusions are summarized in \S \ref{sec:ch4_summary}.  Throughout, we use {\it quasar} to refer to any AGN (regardless of luminosity or radio loudness and including \bl\ objects), and we adopt $H_0= 71$~km s$^{-1}$~Mpc$^{-1}$, $\Omega_m=0.27$, and $\Omega_{\Lambda}=0.73$.

\section{Sample Selection}
\label{sec:ch4_sampsel}

The optically selected \bl\ sample is taken from 8250~deg$^2$ of the SDSS Data Release 7.1 \citep[DR7.1,][]{dr7pap} spectroscopic survey, which includes $1.4 \times 10^6$ spectra.   The SDSS is a multi-institutional effort to image 10$^4$~deg$^2$ of the north Galactic cap in 5 optical filters covering 3800 to 10,000~\AA\ \citep[e.g., see][]{fukugita96,gunn98}, with follow-up moderate resolution spectroscopy ($\lambda/\Delta\lambda\sim1800$) of 10$^6$ galaxies, 10$^5$ quasars, and 10$^5$ unusual stars \citep[e.g., see][]{york00}.  Data are taken with a special purpose 2.5-meter telescope located at Apache Point Observatory \citep[see][]{gunn06}, with astrometric accuracy at the $\sim$100 milli-arcsec level at the survey limit of $r\sim22$ \citep{pier03} and typical photometric precision at brighter magnitudes of approximately 0.02-0.03~mag \citep{ivezic04}; 640 simultaneous spectra are obtained over a 7~deg$^2$ field with a multi-fiber optical spectrograph.  Further technical details can be found in \citet{stoughton02}.  The SDSS DR5 Quasar Catalog \citep{schneider07} contains 77,429 quasars with reliable spectroscopic redshifts.  Our selection process identifies a very small fraction of objects ($<$0.1\%) in the DR5 Quasar Catalog as \bl\ objects.   Although the DR5 Quasar Catalog does not formally include BL Lac objects, some overlap is expected between our BL Lac sample and the quasar catalog, as our \bl\  definition can include some very weak emission line objects. 

 Recent SDSS data releases contain an increasing number of  stellar spectra.  Around 10\% of the spectra in DR7.1 were targeted by the {\it Sloan Extension for Galactic Understanding and Exploration} \citep[SEGUE,][]{yanny09} survey.  SEGUE complements the SDSS Legacy survey (which focused on galaxies and quasars), and it is designed to acquire spectra of 240,000 stars in the Galactic disk and spheroid.  

SDSS data releases typically contain improvements to the reduction pipelines.  Many of the recent spectroscopic improvements pertain to better wavelength calibration (e.g.,  better flat fields, more robust algorithms to calibrate wavelengths from arc lamps, etc.), as required to meet the SEGUE science goals.  Wavelength calibration improvements are typically on the order of $\sim$10~km~s$^{-1}$ and should not  impact \bl\ selection.  However, an important improvement to the spectrophotometric calibration was implemented starting with Data Release 6 \citep[DR6,][]{adelman07}.   The flux scale of SDSS spectra is now tied to point spread function (psf) magnitudes of standard stars, instead of fiber magnitudes as in Data Releases 1--5.  The effect is generally small over the wavelength coverage of the SDSS spectrograph, but flux densities in the far blue end of SDSS spectra can differ by as much as 30\% before and after DR6 \citep[e.g., see Figure~5 of][]{adelman07}.  

The new flux scale can change our measures of blue spectral features, especially the strength of the \ion{Ca}{2}~H/K break ($C$, see Equation~\ref{eq:hkbreak} below).   The \bl\ sample presented here is the first SDSS sample selected post-DR6, so we do not expect {\it all} previously identified SDSS \bl\ objects to make it into this sample (even if we were to apply identical selection approaches.)   However, the majority of previously identified SDSS \bl\ objects should still be recovered.  Previously identified SDSS \bl\ objects that are now rejected because of the improved spectrophotometric calibrations should primarily be those with relatively large H/K breaks (i.e., $0.3 \lesssim C < 0.4$) as measured in their pre-DR6 spectra (but with $C>0.4$ in their DR7.1 spectra.)  We describe the strength of the \ion{Ca}{2}~H/K break, $C$, as the fractional change in continuum blueward and redward of rest-frame 4000~\AA\ \citep[e.g., see][]{landt02}:
\begin{equation}
\label{eq:hkbreak}
C=\frac{\left< f_{\nu,r}\right> - \left<f_{\nu,b}\right>}{\left<f_{\nu,r}\right>} =  \nonumber  0.14 + 0.86\left( \frac{\left< f_{\lambda,r}\right> - \left<f_{\lambda,b}\right>}{\left<f_{\lambda,r}\right>}\right)\! ,
\end{equation}
\\[1mm] \noindent where $\left<f_{\nu,b}\right>$ and $\left<f_{\nu,r}\right>$ refer to the average fluxes per unit frequency just blueward (3750-3950~\AA) and just redward (4050-4250~\AA) of the H/K break, respectively.  The quantities $\left<f _{\lambda,b}\right>$ and $\left<f_{\lambda,r}\right>$ similarly refer to the average fluxes per unit wavelength.  A typical elliptical galaxy has $C\sim0.5$ \citep[e.g., see][]{dressler87,marcha96}; in the context of the standard \bl\ unification model, more highly beamed \bl\ objects have weaker H/K breaks (i.e., smaller $C$ measures, see \citealt{landt02}).

\subsection{Selection Criteria}
\label{sec:ch4_criteria}

We generally require each object's SDSS spectrum to pass the following criteria for inclusion in this \bl\ sample: (1) emission lines show  rest-frame equivalent widths ($REW$) $<5$~\AA; and (2) the \ion{Ca}{2}~H/K~depression has measured $C\le0.4$. These criteria are commonly adopted in the literature and have similarly been applied to assemble previous SDSS \bl\ samples (e.g., \citetalias{collinge05}; \citealt{plotkin08}).

In general, we do not explicitly require a close match to a radio or to an X-ray source for inclusion in this sample.     However, if an object's optical SDSS position matches within 2$''$ to a FIRST radio source, within 10$''$ to an NVSS radio source, and/or within 60$''$ to a RASS X-ray source, then we keep track of its radio and X-ray properties.    We also note that radio and/or X-ray information can be used for some serendipity SDSS spectroscopic target algorithms \citep[e.g., see][]{anderson03}, and we consult radio/X-ray information to avoid  rejecting a small fraction ($\sim$2\%) of  \bl\ candidates with optical colors similar to featureless stars (see \S \ref{sec:ch4_dcwd}) or high-redshift elliptical galaxies (see \S \ref{sec:ch4_lrg}).  Thus, multiwavelength properties play a minor role in our selection technique, but the flux sensitivity of this sample is dominated by the SDSS optical flux limit.

\subsection{SDSS Database Queries}
\label{sec:ch4_query}
 We first query the SDSS spectroscopic database for all objects flagged as having unreliable pipeline redshifts (i.e., {\tt ZSTATUS $=$ `NOT\_MEASURED'} or {\tt `FAILED'}).   This query returned 12,025 spectra, and we refer to it as the `badZ' query.   Next we search the database for spectra with pipeline redshifts {\it not} flagged as {\tt `NOT\_MEASURED'} or {\tt `FAILED'}.   In this query, we retain spectra with pipeline measured $C<0.41$ and pipeline measured $REW<10$~\AA\ for the following spectral lines: Ly$\alpha$, \ion{C}{4}, \ion{C}{3}], \ion{Mg}{2}, [\ion{O}{2}], H$\beta$, [\ion{O}{3}], and H$\alpha$.  This yielded 149,321 spectra, and we refer to it as the `goodZ' query.  Not all pipeline redshifts returned by this query are correct.  However, objects with incorrect redshifts are almost always especially featureless and are retained.  Although SDSS pipeline equivalent width measures are typically reliable, they can sometimes fail for especially weak features; this is the reason  we initially require  {\it REWs} twice as large as our adopted \bl\ classification criteria.  We only extend the $H/K$ break criterion from $C<0.40$ to $C<0.41$ during the database queries because the pipeline H/K break measures are fairly reliable (even for small H/K breaks): independent measurements of the H/K breaks of $\sim$200 objects are nearly identical to the pipeline measures ($\sigma_{rms}\sim0.008$).

As expected, these database queries  recover the majority ($>$90\%) but not all previously known SDSS \bl\ candidates.  For example, 224 of \citetalias{collinge05}'s 240 probable \bl\ candidates are returned by our database queries;  similarly, 465 of the 501 radio-selected \bl\ candidates from \citet{plotkin08} are recovered.  Further details are provided in \S \ref{sec:completeness}.

\subsection{$S/N$~Cut}
\label{sec:ch4_signoise}
We remove especially noisy spectra from our sample by requiring each spectrum to have a signal-to-noise ratio  $S/N>100$ in at least one of 3 spectral regimes centered on 4750, 6250, and 7750~\AA\ with width $\Delta\lambda=500$~\AA.  Each wavelength region typically contains approximately 450, 350, and 300 pixels respectively.  These spectral regions are typically of high quality and fall within the g, r, and i filters respectively, and this $S/N$ constraint roughly corresponds to fiber magnitudes $g < 20.5$, $r < 20.3$, or $i < 19.6$ for a typical spectroscopic plate.  These cuts reduce the number of spectra returned by the `badZ' and `goodZ' queries to 2197 and 140,493 spectra respectively.  More sources are eliminated from the `badZ' query because spectra for which the SDSS pipeline does not find reliable redshifts tend to have poorer $S/N$.

\subsection{Automated Removal of the Majority of Stars and Galaxies}
\label{sec:ch4_poltest}
We are now left with 142,690 spectra, of which, given the area of the sky surveyed, we expect $<$0.5\% to be \bl\ objects.    The bulk of the returned  objects are expected to be stars, especially because about 10\% of the DR7.1 spectra are taken for SEGUE.\footnote{It is possible for \bl\ objects to appear on SEGUE plates, which is why we search these plates despite the increased stellar contamination.}  
Our queries also return quasars with weak emission features, galaxies that underwent recent star formation, and some passive ellipticals that have exceptionally weak $H/K$~breaks.  It is impractical to uniformly visually inspect this many spectra, so we seek additional automated algorithms to reduce the number of spectra that will require manual inspection. 

Following \citetalias{collinge05} (see their \S3.1.2), we make use of the `SpecBS' SDSS spectral classification routines.\footnote{The `SpecBS' routines were written by David~J.~Schlegel.  A brief description is given in \S 4.2 of \citet{adelman07}, and starting with DR6 the `SpecBS' data products are available for download from the SDSS Data Archive Server (DAS, \url{http://www.sdss.org/dr7/access/index.html\#DAS}).}  
The `SpecBS' routines fit each spectrum with various stellar templates (to classify stars) and combinations of eigenspectra (to classify galaxies and quasars).  A second or third-order polynomial is added to the fits to account for deviations of individual spectra from the stellar templates and eigenspectra.  The spectrum's redshift is left as a free parameter, and the stellar template/eigenspectra plus polynomial fit that yields the lowest $\chi^2$ is determined as the best fit.  The source is classified as a star, as a galaxy, or as a quasar according to the preferred stellar template or combination of eigenspectra.  

We use the strength of the polynomial component to remove contaminating spectra.  Normal stars (i.e., with modest absorption features) and normal galaxies should be well fit by stellar templates and galaxy eigenspectra with no need for a strong polynomial component.  Since the odd spectral characteristics of \bl\ objects are caused by non-thermal radiation, \bl\ spectra that are classified as stars or galaxies by the `specBS' routines will require significant polynomial contributions.  There are combinations of quasar eigenspectra that can yield non-thermal spectra with weak emission features without strong polynomial components, so the strength of the requisite polynomial cannot be used to reject spectra classified as quasars by the `specBS' routines.  However, only $\sim$10\% of all SDSS spectra are classified as quasars, so applying a cut based on polynomial strength to even just the objects classified as stars and galaxies will greatly reduce the number of spectra to be examined manually.  

Following \citetalias{collinge05}, we calculate the relative strength of the polynomial component in two spectral regions.  The first region is near the blue end of the SDSS spectrum (3800-4200~\AA), and the second is near the red end (8800-9200~\AA).  For each spectral regime we measure the average total flux density, $\left <f_{tot}\right>$, and we calculate the average flux density from the polynomial, $\left <f_{poly}\right >$.  We then define the polynomial ratio $R=\left <f_{poly}\right >/\left <f_{tot} \right>$.  We reject all objects classified as stars or galaxies for which $R_{blue} < 0.6$ and $R_{red} < 0.5$ (i.e., we reject objects if the polynomial does not account for at least half of the observed flux).  The higher cutoff in the blue is due to the larger number of absorption features in stellar spectra and the presence of the H/K break for some galaxies in this spectral regime.  As a consistency check, we visually inspected a random subset of spectra that failed the above polynomial ratio cut, and those spectra indeed appear to be normal galaxies and stars.  

\begin{figure}
\centering
\includegraphics[scale=0.44]{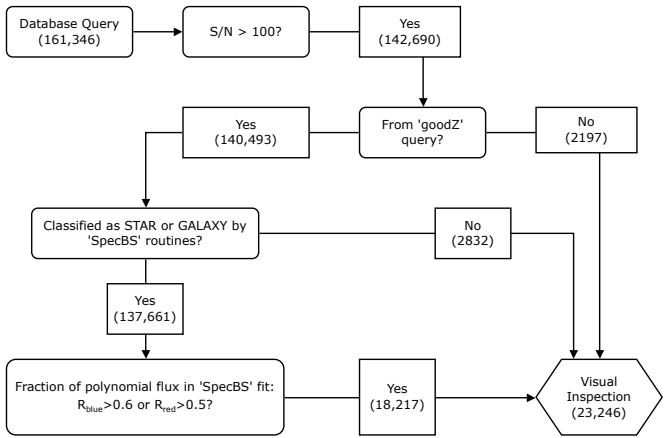}
\caption{Flow chart summarizing the selection process until visual inspection (Sections~\ref{sec:ch4_query} -- \ref{sec:ch4_poltest}).}
\label{fig:ch4_flowchart1}
\end{figure}

We do not perform this polynomial test on the 2197 remaining spectra from the `badZ' query;   we prefer to examine all of those visually since that might lead to serendipitous discoveries of other classes of objects.  We also retain all objects returned by the `goodZ' query that are classified as quasars (2832 spectra) by the `SpecBS' routines.  We apply the polynomial cut to objects classified as stars (81,883 spectra) and galaxies (55,778 spectra).  This cut retains 21,049 spectra from the `goodZ' query (2,832 classified as quasars by `SpecBS', 14,909 stars, and 3,308 galaxies).  The selection process until visual inspection is summarized in Figure~\ref{fig:ch4_flowchart1}.

\section{Visual Inspection}
\label{sec:ch4_visinspec}
We are left with 23,246 spectra to examine visually.  The vast majority of these spectra are stars, and most of them could be rejected through automated routines that search for spectra with common stellar absorption features.  However, stars that would be rejected with such an algorithm can also be very easily and quickly identified by eye.   We thus opted to visually inspect all $\sim$23,000 spectra to expand the opportunity for potential serendipitous discoveries.  We also examine color cutouts of the SDSS images\footnote{\url{http://cas.sdss.org/astrodr7/en/tools/chart/list.asp}} 
to reject objects that are obvious superpositions, blends, or faint low-surface brightness galaxies (which can have weak featured spectra that formally pass our \bl\ selection criteria).  

\begin{figure}
\centering
\includegraphics[scale=0.43]{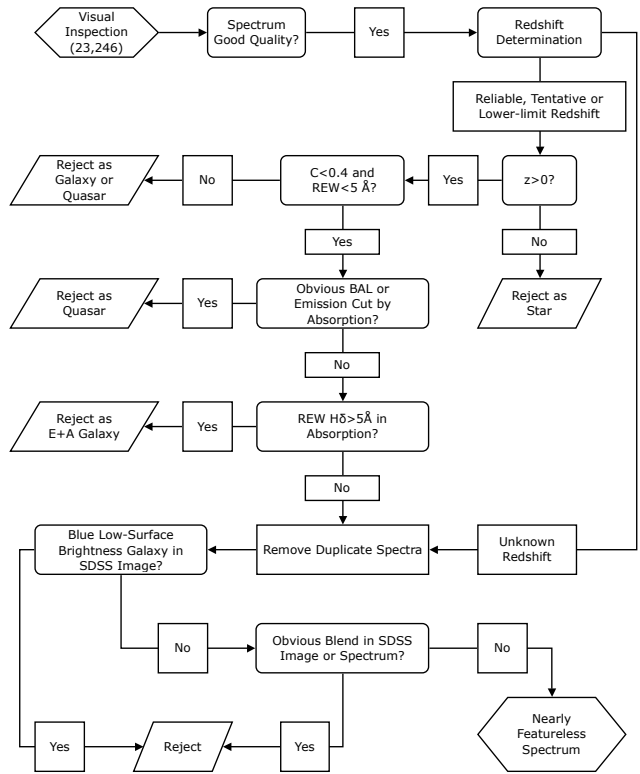}
\caption{Flow chart summarizing the visual inspection process (\S \ref{sec:ch4_visinspec}).}
\label{fig:ch4_flowchart2}
\end{figure}

During visual inspection we attempt to identify a redshift for each spectrum, and we then proceed to classify objects as \bl\ candidates, stars, galaxies, or quasars.  Because we do not consult all SDSS spectral quality flags during our database queries (\S\ref{sec:ch4_query}), a small percentage of  spectra ($\sim$1-2\%) are missing data over portions of the SDSS wavelength coverage.   We reject the small fraction of spectra that are missing substantial portions of data,  and we reject them outright (in addition to the spectra we reject based on their $S/N$). Our visual inspection procedure is outlined below and summarized in Figure~\ref{fig:ch4_flowchart2}.

\subsection{Spectroscopic Redshifts}
We verify the pipeline redshift by searching for at least two spectral features consistent with the pipeline redshift.  If the pipeline redshift is correct, or if it is incorrect but we identify a different redshift consistent with at least two spectral features, then we refer to the redshift as {\it reliable}.   Spectra are also assigned {\it reliable} redshifts if the Ly$\alpha$ forest is the only observed spectral feature.   If the pipeline redshift (or our best estimate) appears correct but the features are especially weak, or the spectrum is too noisy to claim the redshift with high-confidence, then we classify the redshift as {\it tentative}.  A spectrum can also be assigned a {\it tentative} redshift if it shows only a single weak emission feature that we assume to be \ion{Mg}{2}.   A spectrum that shows an absorption doublet consistent with \ion{Mg}{2}~$\lambda$2796,2804 (that might be intrinsic to the source, or caused by intervening material) is said to have a {\it lower limit} redshift.   Note, we require detection of both \ion{Mg}{2} components, so we are not sensitive to intervening absorption systems with broad (i.e., blended) \ion{Mg}{2} absorption.   Finally, a spectrum that is essentially featureless or too noisy to determine a redshift with any confidence is said to have an {\it unknown} redshift.   Objects with {\it unknown} redshift are constrained to have $z<2.2$, or else they would show the Ly$\alpha$ forest in their SDSS spectra.   In the following, we treat all {\it tentative} and {\it lower limit} redshifts as exact.  

\subsection{Spectral Line Measurements}
We measure $C$ (the strength of the \ion{Ca}{2}~H/K~break, see Equation~\ref{eq:hkbreak}) via an automated process for all spectra with reliable, tentative, and lower limit redshifts; spectra with H/K~breaks larger than 40\% are rejected. The $REW$ of the strongest emission line in each spectrum is measured manually with IRAF, and spectra showing any emission line with $REW>5$~\AA\ are rejected.  We fit selected blended lines with multiple Gaussians.  However, we reject spectra that have deblended H$\alpha$+[\ion{N}{2}] components with $REW \leq 5$~\AA\ if they show {\it blended} H$\alpha$+[\ion{N}{2}] with $REW >5$~\AA\ and broad H$\alpha$.  This is under the assumption that deblending underestimates the H$\alpha$ line flux and overestimates the [\ion{N}{2}] line flux.

\subsection{Contaminants Removed During Visual Inspection}
\label{sec:ch4_visinspec_contam}
\subsubsection{Stars}
\label{sec:ch4_stars}
We reject spectra that show obvious stellar absorption features at zero redshift  (e.g., the Balmer series or He for O, B and A stars; \ion{Ca}{2}, \ion{Mg}{1}, and \ion{Na}{1} for F, G and K stars; \ion{Ca}{2}, \ion{Na}{1}, and TiO molecular bands for M stars, etc.).  Approximately 70-75\% of spectra are rejected as obvious stars during visual inspection.  The large stellar contamination rate illustrates one of the disadvantages of not using radio and/or X-ray emission as selection criteria.  Less than 10\% of spectra that required visual inspection in the radio selected sample of \citet{plotkin08} were identified as stars; these stars are likely superpositions with radio sources \citep[e.g., see][]{kimball09}.  Of course, inclusion of such multiwavelength selection criteria likely biases samples toward radio and/or X-ray bright objects.  In any case, the removal of even $\sim$10$^4$ obvious stars is not prohibitively time consuming.

\begin{figure}
\centering
\includegraphics[scale=0.48]{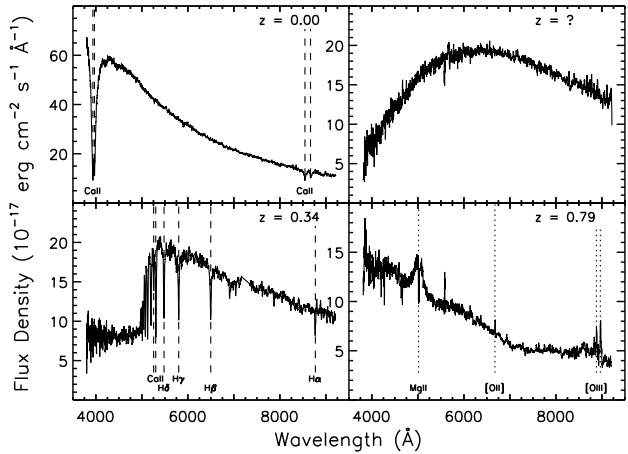}
\caption{Sample spectra rejected during visual inspection.  Selected spectral features are labeled; absorption features are marked with {\it dashed} lines and emission features are marked with {\it dotted} lines.  
{\it Top left:} SDSS J080131.15$+$532900.8, a DZ white dwarf with prominent \ion{Ca}{2} absorption but an otherwise relatively featureless spectrum. 
{\it Top right:} SDSS J122801.59$+$330035.5, a moderately cool DC white dwarf candidate with a relatively featureless spectrum and a blackbody continuum that peaks in the SDSS spectral coverage. 
{\it Bottom left:} SDSS J113338.64$+$041146.2, an E+A galaxy rejected because it shows H$\delta$ with $REW <-5$~\AA\ in absorption. 
{\it Bottom right:} SDSS J144408.82$+$585332.1, a quasar with \ion{Mg}{2} emission cut by absorption.
}
\label{fig:ch4_rejspec1}
\end{figure}

Main sequence stars of all spectral types were identified as contaminants.  We also identified about a dozen cataclysmic variables and a similar number of M-dwarf/white dwarf binaries, all of which were previously known. Thousands of white dwarfs were additionally recovered, including some members of unusual white dwarf subclasses: these include 8 DZ white dwarfs with prominent \ion{Ca}{2} absorption (but otherwise relatively featureless spectra) that are not identified as white dwarfs in SIMBAD;\footnote{The SIMBAD database (\url{http://simbad.u-strasbg.fr/simbad/}) is operated at CDS, Strasbourg, France.} 
we also recovered relatively featureless stars with blackbody continua that peak in the SDSS spectral coverage.  These are likely cool DC white dwarfs; we find 9 that are not cataloged as such in SIMBAD.    The coordinates of these 17 serendipitous discoveries  are listed in Table~\ref{tab:ch4_rarewd}, and sample spectra of two white dwarf candidates are shown in the top row of Figure~\ref{fig:ch4_rejspec1}.  

\begin{deluxetable*}{crrccl}
\tabletypesize{\scriptsize}
\tablewidth{0pt}
\tablecaption{Unusual Subclasses of White Dwarfs \label{tab:ch4_rarewd}}
\tablehead{
	   \colhead{SDSS Name} 
	& \colhead{RA} 
	& \colhead{Dec} 
	& \colhead{$r$} 
	& \colhead{Temperature} 
	& Comment 
\\
	    \colhead{(J2000)} 
	 & \colhead{(J2000)}
	 & \colhead{(J2000)} 
	 & \colhead{(mag)} 
	 & \colhead{(K)}
	 &  \colhead{}
}	
\startdata
080131.15$+$532900.8 &  120.37982  &  53.48358  &  17.49  & \ 9719 & Strong Ca~\Rom{2} \\
082927.85$+$075911.4 &  127.36605  &    7.98652  &  18.69  & 11232 &Strong Ca~\Rom{2} \\
082935.10$+$261746.4 &  127.39627  &  26.29624  &  20.07  & \ 9369 & Strong Ca~\Rom{2} \\
083858.56$+$232252.9 &  129.74401  &  23.38138  &  18.86  & \ 6625 & Strong Ca~\Rom{2}\\
090240.47$+$153556.7 &  135.66864  &  15.59911  &   17.65  & \ 4467  &  Cool DC \\
092430.80$+$312032.3 &  141.12835  &  31.34231  &   17.96  & \ 4653  &  Cool DC \\
094911.91$+$182231.4 &  147.29966  &  18.37541  &  19.98  & 10157 & Strong Ca~\Rom{2} \\
111109.10$+$384856.8 &  167.78795  &  38.81578  &   18.21  & \ 5298  &  Cool DC\\
111316.48$+$285905.6 &  168.31869  &  28.98492  &   17.72  & \ 4726 &  Cool DC\\
120435.37$+$231607.2 &  181.14739  &  23.26868  &   18.58   & \ 4960 &  Cool DC\\
121051.65$+$313659.9 &  182.71523  &  31.61664  &   18.93  & \ 9841 & Strong Ca~\Rom{2} \\
122801.59$+$330035.5 &  187.00664  &  33.00988  &   17.99   & \ 4483 &  Cool DC\\
124805.38$+$141140.3 &  192.02245  &  14.19455  &  19.59  & 13345 & Strong Ca~\Rom{2} \\
125307.45$+$220359.1 &  193.28107  &  22.06644  &   18.40   & \ 5226 &  Cool DC\\
133319.26$+$245046.6 &  203.33029  &  24.84628  &   18.57   & \ 4434 &  Cool DC\\
134156.62$+$133845.3 &  205.48595  &  13.64592  &  19.24  & 11998 & Strong Ca~\Rom{2} \\
162356.37$+$133614.1 &  245.98490   & 13.60392  &   18.66   & \ 4626 &  Cool DC\\
\enddata
\end{deluxetable*}

We fit blackbodies to the spectra of these 17 white dwarfs (excluding \ion{Ca}{2} regions for objects with strong absorption), and we obtain good fits to each star.  The best fit temperature estimates are listed in Table~\ref{tab:ch4_rarewd}.  Note that none of the cool white dwarfs are  at low enough temperature to be confidently considered ``ultra-cool'' ($T<4000$~K), which is also evidenced by the fact that their continua are well-fit by simple blackbodies.  Ultra-cool white dwarfs have significant sources of opacity in their atmospheres from collisions between H$_2$ molecules (or between H$_2$ and He), and simple blackbodies therefore do not fit their spectra well without detailed modeling of their atmospheres \citep[e.g., see][]{farihi05}.

Visual inspection does not remove contamination from especially featureless stars like DC white dwarfs too hot for their continua to peak in the SDSS spectral coverage.  Their removal is discussed in \S \ref{sec:ch4_dcwd}.

\subsubsection{Galaxies}
\label{sec:ch4_gal}
\begin{figure}
\centering
\includegraphics[scale=0.48]{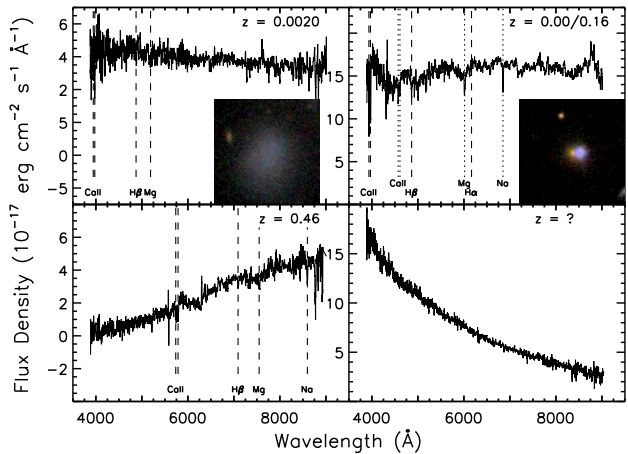}
\caption{More sample rejected spectra, with selected spectral absorption features labeled.  Corresponding SDSS color image cutouts are shown for  the top 2 spectra, as these images were consulted for their removal during visual inspection.  The SDSS images are 50$''$ along each side.  
{\it Top left:} SDSS J111701.05$+$130558.7, rejected during visual inspection as a nearby low-surface brightness galaxy.  
{\it Top right:} SDSS J152036.48$+$133407.3, rejected during visual inspection as a superposition of a (red) galaxy and a (blue) star.  Both components fall within the 3$''$ SDSS optical fiber.  Selected stellar absorption features are labeled and marked with {\it dashed} lines.  Selected absorption features from the galaxy are labeled and marked with {\it dotted} lines.  The galaxy features are all consistent with $z=0.16$. 
{\it Bottom left:} SDSS J212745.21$-$074359.6, rejected post-visual inspection as an LRG (see \S \ref{sec:ch4_lrg}). 
{\it Bottom right:} SDSS J082637.85$+$503357.3, rejected post-visual inspection as a DC white dwarf candidate based on its optical colors (see \S \ref{sec:ch4_dcwd}).  
}
\label{fig:ch4_rejspec2}
\end{figure}

A spectrum is rejected as a galaxy for any of the following reasons (assuming it has a reliable, tentative, or lower limit redshift):
\begin{enumerate}
\item its \ion{Ca}{2}~H/K break is larger than 40\%;
\item any emission feature has $REW > 5$~\AA;
\item it shows H$\delta$ in absorption with $REW < -5$~\AA.  This criterion is designed to remove galaxies that recently underwent a period of star formation but no longer show strong emission features (e.g., E+A galaxies; see \citealt{goto03}).  A sample E+A galaxy spectrum rejected based on H$\delta$ is shown in the bottom left panel of Figure~\ref{fig:ch4_rejspec1};
\item the SDSS cutout image shows an obvious blue, low surface brightness galaxy.  These galaxies, especially because they tend to have poor $S/N$, can possess featureless spectra within our quantitative limits.  A sample spectrum and SDSS image of a rejected low surface brightness galaxy are shown in the top left panel of Figure~\ref{fig:ch4_rejspec2}.
\end{enumerate}
With the above recipe, we reject 5-10\% of the visually inspected spectra as galaxies.  Some passive elliptical galaxies with exceptionally weak H/K~breaks might still contaminate our sample.  However, any such population should be extremely small.  

The above recipe does {\it not} reject a small population of $\sim$80 faint and  extremely red galaxies ($g-r > 1.4$) at $z\gtrsim0.2$.  These galaxies were targeted for SDSS spectroscopy by the Luminous Red Galaxy algorithms \citep[LRG, see][]{eisenstein01}; their removal is discussed in \S\ref{sec:ch4_lrg}.

\subsubsection{Quasars}
Quasars are rejected if any standard emission feature (e.g., Ly$\alpha$, \ion{C}{4}, \ion{C}{3}], \ion{Mg}{2}, H$\beta$, or H$\alpha$) has $REW>5$~\AA.  This may, however, fail to reject quasars that are weak in such standard emission lines but show unusually strong blends of iron emission features.  We also reject objects with obvious broad absorption features, or for which we measure $REW < 5$~\AA\ but the emission feature is clearly cut by absorption (see the bottom right panel of Figure~\ref{fig:ch4_rejspec1}).  Approximately 10\% of the visually inspected spectra are rejected as quasars.  

\subsubsection{A Strange Class of BALQSOs}
We serendipitously recovered/discovered 9 higher-redshift ($1.1 \lesssim z \lesssim 1.9$) objects  that show extreme drop-offs in their continua blueward of rest-frame 2800~\AA, similar to two objects discovered by \citet{hall02}.   These objects are examples of reddened and/or overlapping-trough broad absorption line quasars (BALQSOs).    A similar object (VPMS J1342+2840; $z\sim1.3$) that does not appear in the SDSS spectroscopic database was discovered by \citet{meusinger05}, and two similar low redshift analogs (FBQS 1503+2330; z=0.492 and FBQS 1055+3124; z=0.404) are presented in \citet{white00}.   All previously known objects in this class are radio emitters, but they lack X-ray detections at the RASS sensitivity level.  None of the 9 objects presented here have X-ray detections in RASS, and all but two have radio detections in FIRST.  The two objects lacking FIRST detections are not detected in NVSS either.

\begin{deluxetable*}{crr r@{.}l ccl}
\tabletypesize{\scriptsize}
\tablewidth{0pt}
\tablecaption{Unusual Broad Absorption Line Quasars \label{tab:ch4_oddbal}}
\tablehead{
	   \colhead{SDSS Name} 
	& \colhead{RA} 
	& \colhead{Dec} 
	& \multicolumn{2}{c}{Redshift} 
	& \colhead{$r$} 
	& \colhead{Radio Flux\tablenotemark{a}} 
	& Reference 
\\
	    \colhead{(J2000)} 
	 & \colhead{(J2000)}
	 & \colhead{(J2000)} 
	 & \multicolumn{2}{c}{}  
	 & \colhead{(mag)} 
	 & \colhead{(mJy)} 
	 &  \colhead{}
}	
\startdata
155626.90$+$103457.3 &  239.11212  &  10.58260  &  1 & 889  &  19.45  &  \nodata  &  \\            
134951.94$+$382334.1 &  207.46642  &  38.39281  & 1 & 094 &   19.13  &  1.73  &   3;5 \\                        
160827.08$+$075811.5 &  242.11286  &   7.96988  &  1 & 183 &   16.94  & 13.79  &   3 \\                        
145045.56$+$461504.2 &  222.68984  &  46.25118  & 1 & 877 &   19.54  &  1.51  &   3;4 \\       
220445.26$+$003141.9 &  331.18859  &   0.52832  &  1 & 335 &   17.35  &  3.03  &   1;2;3    \\                        
130941.36$+$112540.1 &  197.42234  &  11.42783  & 1 & 363 &   18.57  &  0.83  &   3;4;5 \\                    
010540.75$-$003314.0 &   16.41980  &     -0.55389  & 1 & 179 &   17.97  &  4.84  &   1;2;3    \\                        
215950.30$+$124718.4 &  329.95960  &  12.78846  &  1 & 514    &   21.03  &  \nodata     &   2;3;5     \\   
161836.09$+$153313.6 &  244.65041  &  15.55378  & 1 & 359\tablenotemark{b} &   18.91  &  2.10  &          \\                        
\enddata

\tablerefs{(1) Identified as an unusual BALQSO in \citet{hall02}; %
(2) cataloged as a BALQSO in \citet{trump06}; %
(3) cataloged as an SDSS quasar in \citet{schneider07}; %
(4) identified as an unusual BALQSO quasar in \citet{plotkin08}; %
(5) cataloged as a BALQSO in \citet{gibson09}. %
}
\tablenotetext{a}{Integrated Flux Density at 1.4 GHz  from FIRST.}
\tablenotetext{b}{Also shows intervening \ion{Mg}{2} and \ion{Fe}{2} absorption at $z=1.156$.}
\end{deluxetable*}

\begin{figure*}
\centering
\includegraphics[scale=0.6]{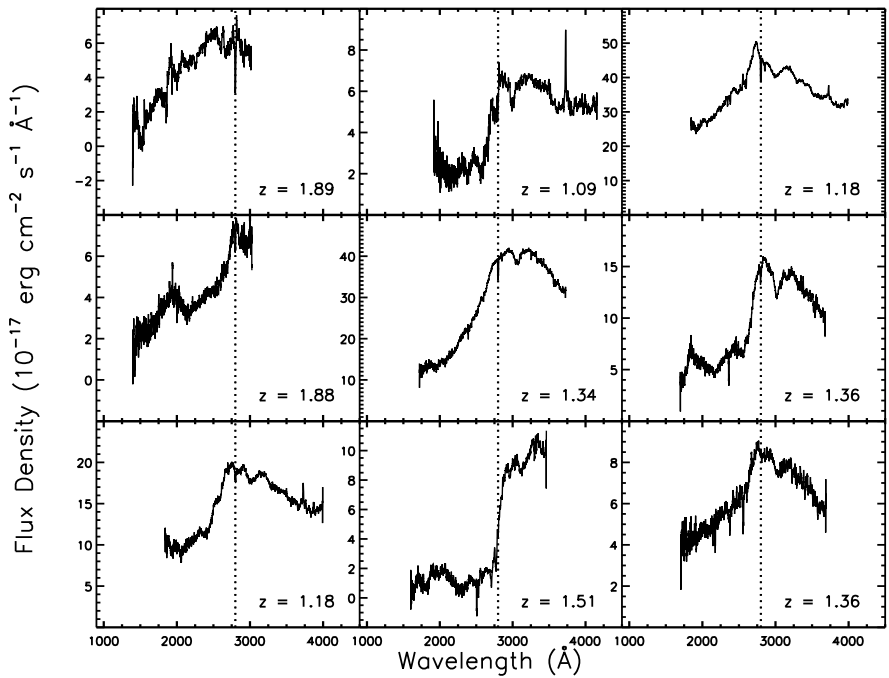}
\caption{SDSS spectra of reddened and/or overlapping-trough BALQSOs, ordered by decreasing strength of \ion{Mg}{2} absorption.   The spectra are shown in their rest-frame, and the location of \ion{Mg}{2} is marked with a vertical dotted line in each panel.}
\label{fig:ch4_oddbal}
\end{figure*}

Basic properties of the 9 recovered objects are given in Table~\ref{tab:ch4_oddbal}, and their spectra are shown in Figure~\ref{fig:ch4_oddbal}.  The sources are presented in order of decreasing strength of \ion{Mg}{2} absorption.  Included in these 9 objects (for completeness) are the two prototypes (SDSS J0105$-$0033 and  SDSS J2204$+$0031) discovered by \citet{hall02}, as well as two similar objects (SDSS J1309$+$1125 and SDSS J1450$+$4615) initially presented in \citet{plotkin08}.  We also note in Table~\ref{tab:ch4_oddbal} whether these objects are cataloged as quasars in  SDSS quasar catalogs, or if they appear in SDSS BALQSO catalogs.  (An object's previous appearance in an SDSS catalog, however, may not indicate that it has been previously associated with the strange objects discovered by \citealt{hall02}).  Each redshift is derived from a \ion{Mg}{2} absorption doublet, except for SDSS J1618+1533.  We assign this object a redshift $z=1.359$ from [\ion{O}{2}] emission and its continuum shape; we interpret \ion{Fe}{2} and \ion{Mg}{2} absorption in its spectrum to be from an intervening  system at $z=1.156$.

The increased sample size of these rare and unusual objects, now covering a broader redshift range, may allow for tighter constraints on their true nature.  We do not include any of these 9 objects as \bl\ candidates.

\subsubsection{ Weak Line Quasars}
The SDSS has discovered a population of $\sim$80 high-redshift quasars with weak emission lines (quantified as Ly$\alpha$+\ion{N}{5} with $REW<10-15$~\AA) at $2.7 \lesssim z \lesssim 5.9$  \citep[e.g., see][]{fan99,anderson01,shemmer06, shemmer09, diamond09}.   Their redshifts can typically only be derived from the onset of the Ly$\alpha$ forest.  While their optical spectra are reminiscent of \bl\ objects, WLQs tend to be weaker radio and X-ray emitters, and they do not show strong polarization.  There does not appear to be a large parent population of radio-loud \bl\ objects at such high-redshifts, and it is unlikely that the SDSS is biased against recovering high-redshift radio and  X-ray bright objects with featureless optical spectra \citep[e.g., see \S 4.1 of ][and \S \ref{sec:ch4_rquiet} of this paper]{shemmer09}.  Thus, WLQs' lack of strong spectral features might not be caused by relativistically boosted jet emission.  However, because WLQs formally pass our \bl\ criteria, we include all high-redshift objects  with $REW<5$~\AA\ without prejudice (but we do identify a subset of objects with weak radio-emission in \S\ref{sec:ch4_alphas}).  

\begin{deluxetable*}{lrrcccc cc}
\tablewidth{0pt}
\tablecaption{New $z>2.2$ Weak Line Quasar Identifications \label{tab:ch4_wlq}}
\tablehead{
	   \colhead{SDSS Name} 
	& \colhead{RA} 
	& \colhead{Dec} 
	& \colhead{Redshift} 
	& \colhead{$z$} 
        & \colhead{$A_z$\tablenotemark{a}}  
        & \colhead{$\alpha_o$\tablenotemark{b}}  
	& \colhead{Radio Flux\tablenotemark{c}} 
	& \colhead{$\alpha_{ro}$\tablenotemark{d}} 
\\
	    \colhead{(J2000)} 
	 & \colhead{(J2000)}
	 & \colhead{(J2000)} 
	 & \colhead{} 
	 & \colhead{(mag)} 
         & \colhead{(mag)} 
         & \colhead{}   
         & \colhead{(mJy)}  
	 &  \colhead{}
}	
\startdata
080906.88$+$172955.2\tablenotemark{e} &   122.27867  &  17.49867   &  2.951 &  18.17  &   0.07 &   0.47  &     $<$0.92 &     $<$0.09 \\
082638.59$+$515233.2\tablenotemark{e} &   126.66080  &  51.87591   &  2.850 &  16.93  &   0.05 &   0.67  &\,\,\,\,1.74 &\,\,\,\,0.04 \\
083304.74$+$415331.3\tablenotemark{e} &   128.26976  &  41.89204   &  2.329 &  18.92  &   0.05 &   0.78  &     $<$0.95 &     $<$0.15 \\
083330.56$+$233909.2\tablenotemark{e} &   128.37736  &  23.65256   &  2.417 &  18.90  &   0.05 &   1.00  &\,\,\,\,2.05 &\,\,\,\,0.20 \\
084249.02$+$235204.7\tablenotemark{e} &   130.70429  &  23.86800   &  3.316 &  18.98  &   0.04 &   1.46  &\,\,\,\,1.55 &\,\,\,\,0.12 \\
084424.24$+$124546.5\tablenotemark{f} &   131.10102  &  12.76294   &  2.466 &  17.60  &   0.07 &   1.03  &\,\,\,\,4.17 &\,\,\,\,0.14 \\
090703.92$+$410748.3\tablenotemark{e} &   136.76634  &  41.13009   &  2.672 &  19.33  &   0.02 &   0.08  &     $<$1.39 &     $<$0.25 \\
092312.75$+$174452.8                                 &   140.80314  &  17.74801   &  2.260 &  18.21  &   0.05 &   1.11  &     $<$0.95 &     $<$0.08 \\
093437.53$+$262232.6\tablenotemark{f} &   143.65639  &  26.37574   &  3.370 &  19.92  &   0.03 &   0.76  &     $<$0.90 &     $<$0.19 \\
101849.78$+$271914.9                                 &   154.70745  &  27.32081   &  2.603 &  18.95  &   0.04 &   1.09  &     $<$1.04 &     $<$0.13 \\
102609.92$+$253651.2                                 &   156.54136  &  25.61425   &  2.318 &  18.75  &   0.03 &   0.54  &\,\,\,\,3.93 &\,\,\,\,0.27 \\
111642.81$+$420324.9\tablenotemark{e} &   169.17840  &  42.05694   &  2.526 &  18.41  &   0.03 &   0.80  &\,\,\,\,1.13 &\,\,\,\,0.12 \\
113747.64$+$391941.5\tablenotemark{e,f}     &   174.44852  &  39.32820   &  2.395 &  18.61  &   0.03 &   1.12  &     $<$0.90 &     $<$0.09 \\
115959.71$+$410152.9\tablenotemark{e}  &   179.99880  &  41.03137   &  2.788 &  17.18  &   0.02 &   0.74  &     $<$0.97 &     $<$0.01 \\
124745.39$+$325147.0\tablenotemark{f}  &   191.93914  &  32.86308   &  2.249 &  18.43  &   0.02 &   0.83  &     $<$0.76 &     $<$0.08 \\
132703.26$+$341321.7\tablenotemark{e}  &   201.76360  &  34.22271   &  2.558 &  18.62  &   0.02 &   0.66  &     $<$0.90 &     $<$0.13 \\
141657.93$+$123431.6\tablenotemark{e} &   214.24138  &  12.57546   &  2.609 &  18.32  &   0.03 &   0.53  &     $<$1.00 &     $<$0.12 \\
144803.36$+$240704.2\tablenotemark{f}  &   222.01404  &  24.11785   &  3.544 &  18.84  &   0.06 &   1.03  &     $<$0.94 &     $<$0.08 \\
213742.25$-$003912.7\tablenotemark{e}   &   324.42605  &  -0.65355 &  2.257 &  19.02  &   0.07 &   0.93  &     $<$0.91 &     $<$0.15 \\
233939.48$-$103539.3\tablenotemark{e,f}     &   354.91454  & -10.59427 &  2.757 &  18.52  &   0.04 &   0.41  &     $<$1.02 &     $<$0.14 \\
\enddata
\tablenotetext{a}{Extinction in the $z$ filter from \citet{schlegel98}.}
\tablenotetext{b}{Optical spectral index ($f_{\nu}\sim\nu^{-\alpha_o}$) measured via fitting a power law to each SDSS spectrum, excluding Ly$\alpha$ and the Ly$\alpha$ forest from each fit (see \S\ref{sec:ch4_specdecomp}).}
\tablenotetext{c}{Integrated Flux Density at 1.4 GHz  from FIRST.}
\tablenotetext{d}{Radio to optical broad-band spectral index referenced at rest-frames 5~GHz and 5000~\AA.  We use the SDSS $z$-band magnitude for this calculation because its wavelength coverage is closest to rest-frame 5000~\AA\ among the SDSS filters.  We adopt $\alpha_{ro}=0.2$ as the approximate boundary between radio-loud and radio-quiet quasars.  See \S\ref{sec:ch4_alphas} for details on the $\alpha_{ro}$ calculation.}
\tablenotetext{e}{Included in the SDSS DR5 Quasar Catalog \citep{schneider07}.}
\tablenotetext{f}{Included as a weak-featured radio-quiet object in Tables~\ref{tab:ch4_empdata_rq} and \ref{tab:ch4_deriveddata_rq} (see \S\ref{optcat}).}
\end{deluxetable*}

Part of the motivation for assembling an optically selected \bl\ sample is to find interesting objects like WLQs that would not be recovered by radio and/or X-ray selection.  In Table~\ref{tab:ch4_wlq} we list \tabnwlq\ serendipitously discovered WLQs (that have reliable redshifts and are new SDSS WLQ identifications, i.e., they do not appear in \citealt{shemmer06,shemmer09} or \citealt{diamond09}) with $REW<10$~\AA\ for Ly$\alpha$+\ion{N}{5}.  For \tabnwlqBL\ of these objects, all emission features show $REW<5$~\AA, and we therefore include those \tabnwlqBL\ objects as \bl\ candidates as well.  The remaining \tabnwlqNotBL\ objects are not included as \bl\ candidates.  All \tabnwlq\ objects in Table~\ref{tab:ch4_wlq} have reliable $z>2.2$, which is where the Ly$\alpha$ forest enters the SDSS spectral coverage.    We also note that the high-redshift nature of WLQs discovered by the SDSS is a selection effect because the Ly$\alpha$ forest is used to determine redshifts.   For example, \citet{diamond09} only searched for WLQs at $z>3$.  There are some lower-redshift radio-weak/quiet objects with weak-featured spectra in the literature: e.g., PG 1407+265 \citep[$z=0.94$,][]{mcdowell95}; PHL 1811 \citep[$z=0.192$,][]{leighly07ii}; 2QZ J215454.3$-$305654 \citep[$z=0.494$,][]{londish04}; SDSS J094533.99+100950.1 \citep[$z=1.66$,][]{hryniewicz09_ph}.  Our \bl\  selection algorithm is also sensitive to such objects (see \S \ref{sec:ch4_rquiet}).

 \citet{smith07} found that optically selected SDSS \bl\ candidates with $z>1$ do not show strong polarization ($<3\%$), suggesting that some $z > 1$ \bl\ candidates might not have continua dominated by beamed synchrotron radiation.    Perhaps $z>1$ \bl\ candidates are actually a mix of \bl\ objects and lower-redshift WLQs.  The resolution of this issue is unclear with extant data; additional observations (such as variability and polarization monitoring, as well as deeper radio/X-ray imaging for objects lacking radio/X-ray detections) are needed.  Thus, we include all $z>1$ objects that formally pass our \bl\ criteria as \bl\ candidates, but we recognize that some of these objects might rather constitute an equally fascinating population of lower-redshift analogs to WLQs.   This issue is discussed in more detail in \S \ref{sec:ch4_rquiet}.

\subsubsection{Duplicate Spectra and Superpositions}
A small fraction of SDSS objects ($<$1\%) have repeat spectroscopic observations.  Multiply-observed sources only need to pass our selection criteria at a single epoch for inclusion in our sample.  If multiple spectra of the same object are classified as \bl s, then we only include one (randomly selected) spectrum in our final sample.   

The combination of flux contributed by two distinct superposed sources can sometimes act to make the observed spectrum appear featureless.   Around 1-2\% of the manually inspected spectra are random superpositions.   Superpositions are identified by sets of spectral features that are consistent with two different redshifts; some superpositions are also identified as  two objects that fall within the 3$''$ SDSS optical fibers, as judged by examining their SDSS images.    The vast majority of superpositions are foreground stars over galaxies, but we do see a small number of galaxy/galaxy, galaxy/quasar, and star/quasar pairs.  An example of a spectrum of a galaxy superposed with a foreground star and a thumbnail of its SDSS image is shown in the top right panel of Figure~\ref{fig:ch4_rejspec2}.    We constructed our database queries to initially retain such superpositions, because otherwise we might reject real \bl\ objects that appear as blue point sources at the centers of elliptical galaxies.  

\section{Removal of Contaminants That Survive Visual Inspection}
\label{sec:ch4_postvisinspec}

\begin{figure}
\centering
\includegraphics[scale=0.45]{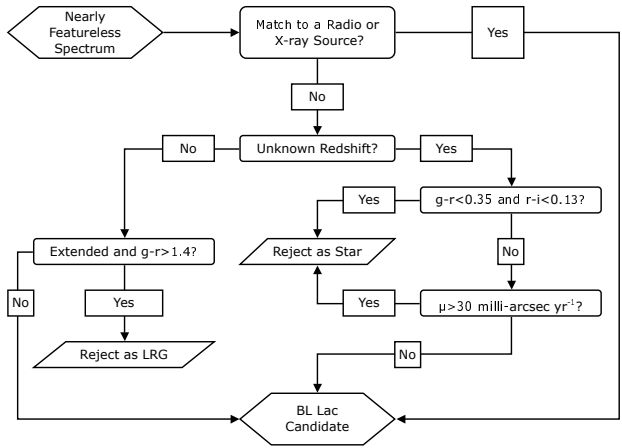}
\caption{Flow chart summarizing the removal of stars and galaxies post-visual inspection (\S \ref{sec:ch4_postvisinspec}).}
\label{fig:ch4_flowchart3}
\end{figure}

Visual inspection reduced our list of nearly featureless spectra to \noptfeat\ objects.  However, there are still surviving contaminants, including DC white dwarfs and high-redshift LRGs.  Their removal is discussed below and summarized in Figure~\ref{fig:ch4_flowchart3}.

\subsection{Removing Featureless White Dwarfs}
\label{sec:ch4_dcwd}
\begin{figure*}
\centering
\includegraphics[scale=0.65]{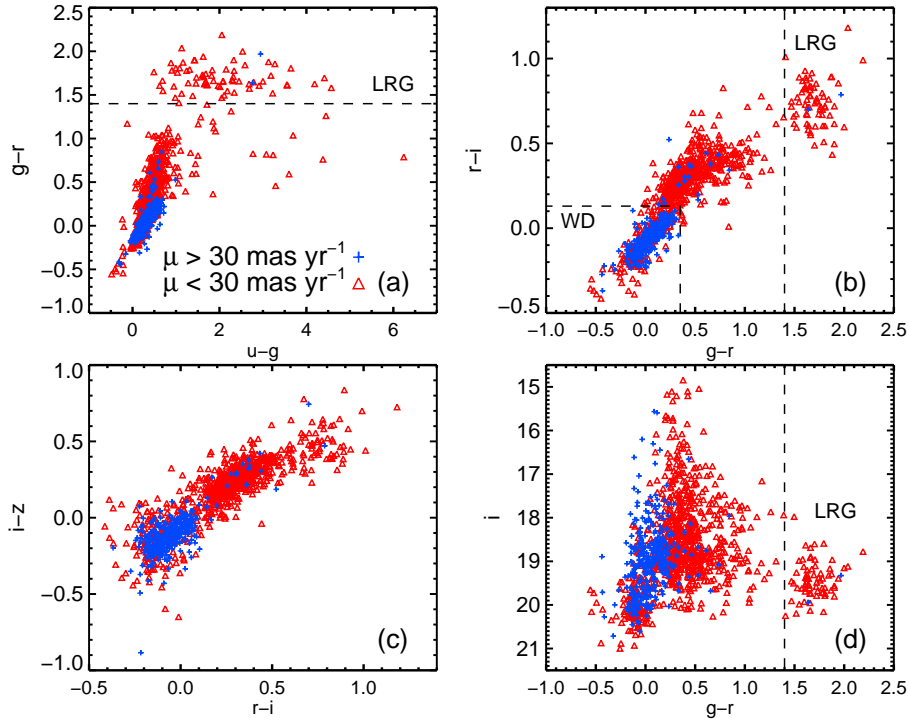}
\caption{Optical colors for 1222 nearly featureless spectra with exactly one match in the proper motion catalog and surviving visual inspection.  Objects with $\mu>30$~milli-arcsec~yr$^{-1}$ are denoted as blue plus signs, objects with $\mu<30$~milli-arcsec~yr$^{-1}$ are shown as red triangles.  Dashed lines on relevant panels indicate color cuts applied to remove DC white dwarfs (WD), $(g-r,r-i)<(0.35,0.13)$, and high-redshift Luminous Red Galaxies (LRG), $g-r > 1.4$.  (a) $g-r$ vs.\ $u-g$; (b) $r-i$ vs.\ $g-r$; (c) $i-z$ vs.\ $r-i$; (d) $i$ vs.\ $g-r$.  All panels use psf magnitudes not corrected for Galactic extinction. Not shown are $\sim$90 nearly featureless spectra lacking proper motion information (or with more than one match in the proper motion catalog.)}
\label{fig:ch4_colpm}
\end{figure*}

We use proper motions from the SDSS+USNO-B proper motion catalog \citep{munn04} to verify the result from \citetalias{collinge05} that high proper motion objects with nearly featureless spectra tend to populate bluer regions of optical color space than \bl\ objects.  These contaminants are most likely DC white dwarfs.  Figure~\ref{fig:ch4_colpm} shows optical color-color plots for 1222 surviving objects with nearly featureless spectra also in the SDSS+USNO-B catalog \citep{munn04}.  We require each object to match to only a single USNO-B source with a 1$''$ error circle centered on the SDSS position.   Blue plus signs denote objects with measured proper motion $\mu > 30$~milli-arcsec~yr$^{-1}$, and red triangles have $\mu<30$~milli-arcsec~yr$^{-1}$.  We choose $\mu=30$~milli-arcsec~yr$^{-1}$ as the division between large and small proper motions because only 0.8\% of spectroscopically confirmed SDSS quasars in \citet{schneider07} with exactly one match to a USNO-B source have $\mu>30$~milli-arcsec~yr$^{-1}$.  

Figure~\ref{fig:ch4_colpm} indeed confirms that objects with large proper motion tend to have bluer colors.  We adopt the color cut of \citetalias{collinge05}, $g-r < 0.35$ and $r-i < 0.13$,  to remove nearly featureless stars.  We refer to this blue color space as the $gri$ box, and this cut removes 500 objects.  Not all stars will have large measured proper motions (and $\sim$7\% of our nearly featureless spectra do not have proper motion measures), so we prefer to reject objects based on color rather than measured proper motion.   It is possible for some \bl\ objects to populate the $gri$ box, so we retain spectra with blue colors that have reliable, tentative, or lower limit redshifts and/or that match to a radio and/or X-ray source.   We retain 61 such objects as \bl\ candidates.

As stated above, objects inside the $gri$ box with {\it unknown} redshifts are rejected as likely DC white dwarfs.  However,  there are five objects inside the $gri$ box with {\it unknown} redshifts that also match to a radio and/or X-ray source, and we retain this small number of objects as \bl\ candidates because of their multiwavelength properties.   Approximately 84\% of our final \bl\ candidates (see \S \ref{sec:ch4_finalsamp}) match to a FIRST/NVSS radio source or to a RASS X-ray source.  If a similar percentage of very blue \bl\ objects have radio or X-ray emission above the relevant multiwavelength survey flux limits, then we expect our color cut aimed at removing DC white dwarfs to only reject $\sim$1 \bl\ object from our final sample.

In summary, we remove $\sim$\noptrejdc\ nearly featureless spectra (probably DC white dwarfs) that: 
\begin{enumerate}
\item fall within the $gri$ box; 
\item have unknown redshifts; 
\item and do not match to a radio and/or X-ray source.
\end{enumerate}

\subsection{Removing Stellar Contaminants Outside the $gri$ Box}
\label{sec:ch4_bigpm}
A very small number of objects outside the $gri$ box have measured $\mu > 30$~milli-arcsec~yr$^{-1}$.   We reject those objects if they:
\begin{enumerate}
\item have $\mu > 30$~milli-arcsec~yr$^{-1}$ and match to exactly one source in the SDSS+USNO-B proper motion catalog;
\item have an unknown redshift;
\item and do not match to a radio and/or X-ray source.
\end{enumerate}
The above criteria remove only 4 objects that would not have been removed via other cuts.  This is our only cut based solely on proper motion.  Since less than 1\% of spectroscopically confirmed SDSS quasars show measured proper motions larger than 30~milli-arcsec~yr$^{-1}$, with this cut we risk losing $<0.04$ \bl\ objects.  

\subsection{Removing High-Redshift Luminous Red Galaxies}
\label{sec:ch4_lrg}
There is a small population of higher-redshift objects ($z \gtrsim 0.2$) targeted for spectroscopy as LRGs that formally pass our \bl\ criteria after visual inspection.  These objects occupy a distinct region of optical color space, and we separate them phenomenologically as having $g-r > 1.4$. This division is drawn as a dashed line on the relevant panels in Figure~\ref{fig:ch4_colpm}, where their separation from the rest of our nearly featureless spectra can be seen.   These objects formally pass our $S/N$ cut, but they tend to do so only at the redder wavelength regions that we consider in \S \ref{sec:ch4_signoise}.  This is because the \ion{Ca}{2}~H/K~break crosses from the $g$ to the $r$ filter near $z\sim 0.4$.  These spectra have acceptable $S/N$ redward of rest-frame 4000~\AA, but they have relatively low flux blueward of rest-frame 4000~\AA;  their H/K~break measurements are therefore relatively noisy.   For example, the average measured uncertainty  for the H/K~break for these nearly featureless objects with $g-r>1.4$ is $\sigma_C \sim 0.04$, compared to $\sigma_C \sim 0.01$ for all nearly featureless spectra with $g-r < 1.4$.    We believe the majority of these objects are normal LRGs (i.e., with $C\sim0.5$), but because of their noisier H/K~break measures, we find an unexpectedly large number of them to have $C<0.4$.  

We take precautions to not remove red \bl\ objects with $g-r>1.4$ from our sample if they exist.  We thus only remove objects as high-redshift LRGs if they show all of the following:
\begin{enumerate}
\item $g-r>1.4$;
\item their SDSS image morphology is extended (i.e., the SDSS morphology parameter {\tt TYPE}~=~3);
\item and they do not match to a radio and/or X-ray source.
\end{enumerate}
We exclude \noptrejlrg\ spectra that fulfill the above requirements, leaving 17 objects with $g-r>1.4$ that remain as \bl\ candidates.  We visually examined the SDSS spectra and images of these \noptrejlrg\ rejected spectra, and we confirm they indeed look like low $S/N$, extended LRGs.

\section{The Optically Selected Sample}
\label{sec:ch4_finalsamp}
The final sample includes \noptall\ objects.  We subdivide these into \noptrl\ radio-loud \bl\ candidates and \noptrq\ radio-quiet objects with nearly featureless spectra (defined by having $\alpha_{ro}<0.2$, see \S \ref{sec:ch4_alphas}).  The radio-quiet objects may be a mix of \bl\ objects on the radio-faint tail of the \bl\ distribution and a distinct alternate class (see discussion below) of exotic AGN with extremely featureless spectra.  We further subdivide each subset into high-confidence (`H') and low-confidence (`L').   We classify an object as low-confidence if it is unclear where to define the continuum near an emission feature, and that emission feature can have measured $REW$ larger or smaller than 5~\AA\ depending on the continuum used.  We recover \noptrlHigh\ and \noptrlLow\ radio-loud high- and low-confidence \bl\ candidates respectively; and we find \noptrqHigh\ and \noptrqLow\ high- and low-confidence radio-quiet objects with nearly featureless spectra.  Sample spectra for 4 (radio-loud) \bl\ candidates are shown in Figure~{\ref{fig:ch4_blspec}, and basic sample demographics are summarized in Table~\ref{tab:ch4_demo}.   We compare the optically selected sample to previously published SDSS \bl\ catalogs (i.e., \citealt{anderson03}; \citetalias{collinge05}; \citealt{anderson07}; and \citealt{plotkin08}), and we also examine each source's classification in NED.\footnote{The NASA/IPAC Extragalactic Database (NED) is operated by the Jet Propulsion Laboratory, California Institute of Technology, under contract with the National Aeronautics and Space Administration.}  
From this, we estimate that $\sim$35\% of the optically selected \bl\ candidates are new \bl\ identifications, and $\sim$75\% of all of our optically selected \bl\ candidates were initially discovered by the SDSS (either by this or previous studies).   

 We loosely refer to the radio-quiet subset as \bl\ candidates until they are discussed in more detail in Sections \ref{sec:ch4_alphas} and \ref{sec:ch4_rquiet}, where we compare them to WLQs.   Distinguishing these objects from WLQs is not trivial, especially because \noptRquietBigZ\ of the \noptrq\ radio-quiet \bl\ candidates have  $z_{spec}>2.2$ and pass the criteria to alternatively be classified as WLQs.  Note,  \tabnwlqBL\ of those \noptRquietBigZ\ radio-quiet objects also appear in Table~\ref{tab:ch4_wlq} as new WLQ discoveries.  The other \tabnwlqNotBL\ new WLQ identifications in Table~\ref{tab:ch4_wlq} have emission lines too strong to pass our stricter \bl\ $REW<5$~\AA\ criterion (and therefore do not appear in the optically selected \bl\ sample).  

\begin{figure}
\centering
\includegraphics[scale=0.48]{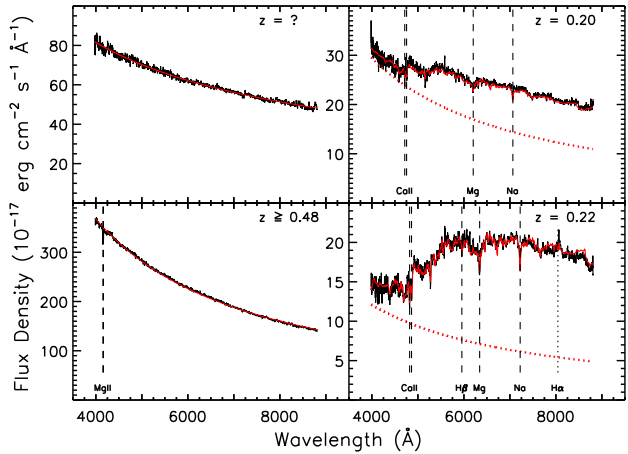}
\caption{Sample spectra of 4 \bl\ candidates, with selected spectral features labeled.  Absorption features are marked with dashed lines, and emission features are marked with dotted lines.   All 4 objects match to FIRST/NVSS radio sources and RASS X-ray sources.   The spectral fits from the host galaxy decomposition described in \S \ref{sec:ch4_specdecomp} are overplotted as red lines.  If there is significant host galaxy contamination, then the dotted red lines illustrate just the power law flux from the AGN.  
{\it Top left:} SDSS J092915.43$+$501336.1, a high-confidence \bl\ candidate with a spectrum too featureless to derive a spectroscopic redshift.  
{\it Top right:} SDSS J084712.93$+$113350.2, a high-confidence \bl\ candidate with reliable $z=0.198$ derived from host galaxy spectral features.  
{\it Bottom left:} SDSS J124312.73$+$362743.9, a high-confidence \bl\ candidate with $z\geq0.485$ derived from intervening \ion{Mg}{2} absorption. 
{\it Bottom right:} SDSS J164419.97$+$454644.3, a low-confidence \bl\ candidate with $z=0.225$ derived from host galaxy spectral features.  This object is classified as low-confidence because it shows broad $H\alpha$ that can have measured $REW>5$~\AA\ depending on the continuum adopted during the measurement.  
}
\label{fig:ch4_blspec}
\end{figure}

\begin{deluxetable}{l@{\hspace{8mm}}  cc c@{\hspace{6mm}} cc}
\tablewidth{0pt}
\tablecaption{Optically Selected Sample Demographics}
\tablehead{
	\nodata & 
	\multicolumn{2}{c}{Radio-Loud}  & & 
	\multicolumn{2}{c}{Radio-Quiet}
\\		
	  \nodata  
	& \colhead{H\tablenotemark{a}} 
	& \colhead{L\tablenotemark{b}} & 
	& \colhead{H\tablenotemark{a}} 
	& \colhead{L\tablenotemark{b}}
}	
\startdata
    Total Number		&  613 & 24 & & 56    &  30 \\
     \hline \hline \\
    Reliable $z_{spec}$      & 216 & 15   &   & 22  &  16  \\
    Tentative $z_{spec}$     & 85 & 9       &   & 22  &  14  \\
    Lower Limit $z_{spec}$ & 42   & 0     &   &    7  &    0 \\
    Unknown $z_{spec}$     & 270 & 0    &   &     5 &  0 \\
    Radio Matches                 & 571 & 19 &   &    9 &  0 \\
    X-ray Matches                  & 281 &   8  &   &    2    &  3 \\
    LBL\tablenotemark{c} 	 & 188 & 4    &    &   0    &  0  \\
    HBL\tablenotemark{d} 	 & 425 & 20  &    &   56  &  30 \\ 
\enddata
\tablenotetext{a}{High-confidence \bl\ candidate}
\tablenotetext{b}{Low-confidence \bl\ candidate}
\tablenotetext{c}{Low-energy peaked \bl\ objects}
\tablenotetext{d}{High-energy peaked \bl\ objects}
\label{tab:ch4_demo}
\end{deluxetable}

\subsection{Tests on the Completeness of Our Recovery Algorithm}
\label{sec:completeness}
We test our recovery algorithm by correlating our sample to the 226 probable \bl\ candidates recovered by \citetalias{collinge05} that also appear in the DR7.1 database\footnote{14 of \citetalias{collinge05}'s probable \bl\ candidates do not appear in the DR7.1 database.} 
 (see \S \ref{sec:ch4_query}): we recover all but 28 of these objects.  Essentially all of the missing \citetalias{collinge05}\ objects have spectral features near the quantitative \bl\ criteria limits.  Although our recovery approach is similar to \citetalias{collinge05}, since we use different spectral reductions it is not surprising that some ``borderline'' objects do not pass the spectral criteria here.  Among the 28 missing objects, 2 were rejected by our database queries; we also rejected 11 objects because they showed poor $S/N$ (\S \ref{sec:ch4_signoise}), they failed the polynomial strength test (\S \ref{sec:ch4_poltest}), or they were deemed to have poor quality spectra/photometry during visual inspection; 10 objects were rejected during visual inspection because they showed emission features with $REW>5$~\AA\ in their DR7 spectral reductions; and the remaining 5 objects were rejected as stars (either during visual inspection or after applying the color cuts described in \S \ref{sec:ch4_postvisinspec}).

We perform a similar quality assurance test against the 500 objects in the DR7.1 database that appear in the \citet{plotkin08} radio-selected sample.\footnote{1 object does not appear in the DR7.1 database.} 
  Our optical sample does not include 95 of those 500 radio-selected objects, of which 35 were rejected by our database queries.  Again, the majority of missing objects tend to have spectral features near our quantitative limits, and we measure their features to be too strong in the DR7 spectral reductions (but their features pass our spectral criteria in earlier data release spectral reductions).  An exception is a small population of 5 spectra that we noticed to be superpositions of two sources in their DR7 spectra.  Of the other 55 missing spectra, 33 showed poor $S/N$ (\S \ref{sec:ch4_signoise}), failed the polynomial strength test (\S \ref{sec:ch4_poltest}), or were deemed to have poor quality spectra during visual inspection;  21 objects were rejected because they showed emission features with $REW>5$~\AA\ in their DR7 spectra; and 1 spectrum was classified as a star during visual inspection.  

Given the post-DR6 improvements to the spectrophotometric pipeline, the fraction of previously identified SDSS \bl\ candidates ($>$80\%) recovered by this new optically selected sample is satisfactory.  As expected, the missing $\sim$20\% are primarily ``borderline'' \bl\ candidates. 

\subsection{Spectral Decomposition: Estimating AGN Spectral Indices, Redshift Limits, AGN Component Fluxes, and Host Galaxy Fluxes}
\label{sec:ch4_specdecomp}
We use each object's SDSS spectrum to measure the AGN's optical spectral index, assuming the AGN flux follows a power law.  If the host galaxy contributes significantly to the total observed flux, then we perform a spectral decomposition of the host galaxy from the SDSS spectrum.  This decomposition additionally allows estimates of optical fluxes (or limits) for the host galaxy and AGN separately.  

If the \bl\ object has an unknown redshift, then (by definition) its spectrum does not show significant contamination from the host galaxy.   For the \noptallNoz\ \bl\ candidates with unknown redshifts, we fit  their spectra (via Levenberg-Marquardt least-squares regression) with the following power law:
\begin{equation}
f_{\lambda,total}(\lambda) = f_{\lambda_0} \left ( \frac{\lambda}{\lambda_0}\right )^{-\alpha_\lambda}.
\label{eq:ch4_specind}
\end{equation}
  $f_{\lambda,total}(\lambda)$ refers to the observed flux density per unit wavelength, and $f_{\lambda_0}$ is the flux density at reference wavelength $\lambda_0=6165$~\AA\ (the effective wavelength of the SDSS $r$ filter).  There are typically $\sim$3800 degrees of freedom (dof), and we estimate the goodness of this power law (pl) fit with the reduced $\chi^2_{dof,pl}$.  
   
If a spectrum has a measured reliable, tentative, or lower limit redshift, then  the \bl\ host galaxy contribution to the observed spectrum might be important.  In addition to fitting Equation~{\ref{eq:ch4_specind} (which has 2 free parameters, the spectral index $\alpha_{\lambda}$ and the flux density normalization $f_{\lambda_0}$), we also perform a power law fit including the contaminating flux from the host galaxy, which we assume to be a normal elliptical\footnote{Normal in every sense other than its active nucleus.} 
\citep[see][]{urry00_hstii}.   Thus, we find the best fit to the following function:
\begin{equation}
f_{\lambda,total}(\lambda) = C_1f_{\lambda,ell} (\lambda,z)+ f_{\lambda_0}\left(\frac{\lambda}{\lambda_0}\right)^{-\alpha_{\lambda}}.
\label{eq:ch4_decomp}
\end{equation}
$f_{\lambda,total}(\lambda)$ is again the observed flux density, and $f_{\lambda,ell}(\lambda,z)$ is the flux density from the elliptical galaxy (at the observed redshift).  We use the elliptical galaxy template from \citet{mannucci01} to model the host galaxy flux density.   This fit has 3 free parameters, the elliptical galaxy flux density normalization ($C_1$), the power law flux density normalization ($f_{\lambda_0}$), and the spectral index of the active  nucleus ($\alpha_{\lambda}$).  We constrain $C_1$ and $f_{\lambda_0}$ to always be positive, and we estimate the goodness of the host galaxy decomposition with the reduced $\chi^2_{dof, decomp}$.   We exclude the Ly$\alpha$ forest from the fits (for spectra with $z>2.2$).

Equation~\ref{eq:ch4_decomp} produces much better fits than Equation~\ref{eq:ch4_specind} for objects with redshifts derived from host galaxy spectral features.  When the host galaxy becomes negligible (as is the case for higher-redshift and/or highly beamed \bl\ candidates), Equation~\ref{eq:ch4_decomp} is identical to Equation~\ref{eq:ch4_specind}.  We use the spectral index returned by the fit with the smaller $\chi^2_{dof}$, in order to minimize the requisite number of free parameters.  We adopt the value $\alpha_{\nu}=2-\alpha_{\lambda}$ returned by whichever fit is preferred.\footnote{We report $\alpha_{\nu}$ rather than $\alpha_{\lambda}$ for consistency (and easy comparison) with the bulk of the literature.}  
 The spectral fits are plotted as red lines over the sample \bl\ spectra in Figure~\ref{fig:ch4_blspec}.  For the cases where the fit including the host galaxy decomposition was preferred, the power law spectrum contributed by just the AGN component is shown as a dotted red line.

The distribution of measured spectral indices is shown in Figure~\ref{fig:ch4_fluxes}a.  The average spectral index is $\left <\alpha_{\nu} \right >=1.15$ (median $\alpha_{\nu}=1.06)$, with a dispersion of 0.69.  The spread in the ensemble's  $\alpha_{\nu}$ distribution is much larger than the  typical measurement uncertainties of $\sigma_{\alpha_{\nu}}=0.01-0.03$.  It is possible that some of the scatter in $\sigma_{\alpha_{\nu}}$ can also be attributed to using the same host galaxy template for each object.   This of course only affects the \bl\ candidates that preferred the spectral decomposition over fitting just a power law fit (around 40\% of the sample).    The dispersion in $\alpha_{\nu}$ for just the objects that preferred the power law fit is similarly large.   Variability also contributes to the observed scatter in $\alpha_{\nu}$.   There are 46 \bl\ candidates with multiple epochs of SDSS spectroscopy (about two-thirds have two epochs, most of the others have three epochs.)  We measured $\alpha_{\nu}$ for each duplicate spectrum (decomposing the host galaxy flux when necessary), and we calculated the difference between the largest and smallest measured $\alpha_{\nu}$ for each object.  The root mean square of the differences in $\alpha_{\nu}$ is $\sim$0.3, which is significant, but unlikely large enough to account for the {\it all} of the observed dispersion in the entire \bl\ sample.   Thus, we conclude that the relatively large dispersion indicates that \bl\ objects do not have universal spectral shapes; the large scatter is probably due to a combination of different amounts of relativistic beaming among objects, variability, and innate physical differences (such as black hole mass, accretion rate, etc).    

 \begin{figure}
\centering
\includegraphics[scale=0.5]{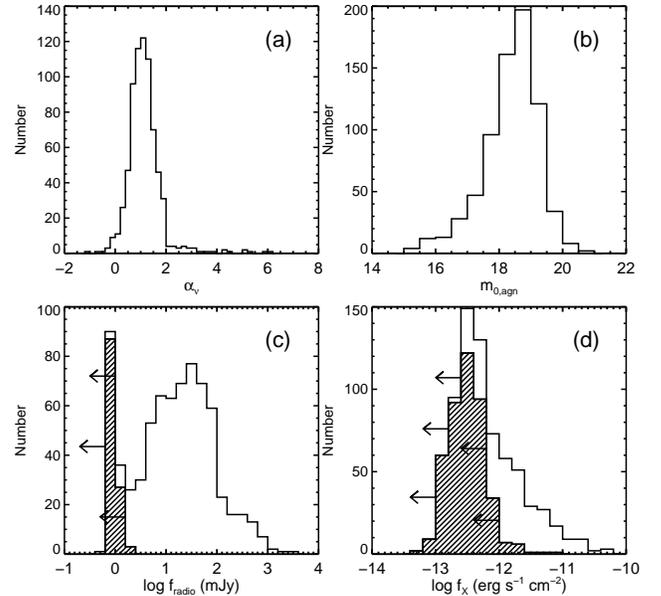}
\caption{AGN component optical spectral indices, and optical, radio, and X-ray flux distributions for \bl\ candidates.  
(a) The optical spectral index $\alpha_{\nu}$ for the decomposed AGN flux.   The width of the histogram indicates that the shape of the AGN's optical  spectrum is not universal among \bl\ objects.   
(b) Extinction corrected optical magnitudes of the decomposed AGN component flux in the SDSS filter closest to 5000~\AA\ rest-frame.  
(c) Logarithm of integrated radio flux densities at 1.4~GHz.  When available, flux densities are taken from FIRST, otherwise flux densities are taken from NVSS.  The hatched histogram shows radio flux density upper limits from FIRST for objects lacking radio detections in both surveys.    
(d)  Logarithm of broad-band X-ray fluxes (corrected for Galactic absorption) from 0.1-2.4 keV.  The hatched histogram shows X-ray flux upper limits for objects lacking X-ray detections in RASS.
}
\label{fig:ch4_fluxes}
\end{figure}

\subsubsection{Estimating Redshift Limits From Host Galaxy Contamination in SDSS Spectra}
\label{sec:ch4_zlim}
\bl\ objects live (perhaps exclusively) in giant elliptical galaxies, and  their hosts tend to have similar luminosities within $\pm$1~mag \citep{scarpa00_hsti,urry00_hstii}.   \citet{sbarufatti05} note that the dispersion in host galaxy absolute magnitudes is narrow enough to treat \bl\ host galaxies as standard candles, allowing the derivation of ``imaging'' redshifts if the host galaxy can be spatially resolved from the AGN.  Using a subset of 64 objects with {\it Hubble Space Telescope (HST)} resolved host galaxies and known spectroscopic redshifts, they show that host galaxy apparent magnitudes are tightly constrained with redshift on a Hubble diagram if they have absolute magnitudes $M_R=-22.9\pm0.5.$\footnote{\citet{sbarufatti05} use $H_0=71$~km~s$^{-1}$~Mpc$^{-1}$, $\Omega_{\Lambda}=0.7$, and $\Omega_m=0.3$, similar to our adopted cosmology.}  
 Their standard candle algorithm is calibrated out to $z\sim0.7$, and it provides ``imaging'' redshifts accurate to $\sigma_{z}\sim\pm0.05$ if a host galaxy is resolved; their technique can also be used to estimate redshift lower limits in the cases of host galaxy non-detections.

We use the result from \citet{sbarufatti05} to estimate redshift lower limits for objects in our SDSS sample that lack spectroscopic redshifts.\footnote{These objects have redshift upper limits $z<2.2$, or else they would show the Ly$\alpha$ forest.} 
  However, instead of attempting to spatially resolve host galaxies (which becomes more challenging with increasing redshift, especially with ground-based telescopes), we rather take advantage of each object's high-quality spectrum.    With extant SDSS data, a spectral decomposition is likely more sensitive than a spatial decomposition: spectral decomposition is primarily limited by the spectral resolution (and collecting area) of the telescope, while spatial decomposition limitations are dominated by the spatial resolution.   For example, only $\sim$200 SDSS \bl\ candidates appear extended in their SDSS images, and essentially all of these objects already have spectroscopic redshifts (and they show host galaxy contamination in their spectra.)   However, some objects with unresolved host galaxies in their SDSS images still show host galaxy features in their SDSS spectra.

Our goal is to estimate an $R$-band apparent magnitude limit (in the Cousins filter system) for each host galaxy.  Equipped with these flux estimates, we can then  translate apparent magnitudes into redshift limits by assuming the hosts are standard candles.   In the previous subsection, approximately 200 high-confidence \bl\ candidates with reliable redshifts preferred the host galaxy decomposition fit (Equation~\ref{eq:ch4_decomp}) over just a power law fit (Equation~\ref{eq:ch4_specind}).  Over 90\% of these objects have decomposed AGN flux densities at most 5 times larger than their host galaxy flux densities at 6581~\AA\ \citep[the effective wavelength of the $R$ filter;][]{fukugita96}.  Therefore, for the \noptallNoz\ \bl\ candidates lacking spectroscopic redshifts, we assume their host galaxy flux densities are at least 5 times fainter than the observed flux densities at 6581~\AA.  Otherwise we would have seen host galaxy features in their spectra during visual inspection and assigned them spectroscopic redshifts.  

We place upper limits to the host galaxy flux densities per unit frequency at the effective wavelength of the $R$ filter ($\lambda_{eff,R}=6581$~\AA) as:
\begin{equation}
f_{\nu,hg}(z) \leq 0.2 \times \left[\left(\frac{\lambda_{eff,R}^2}{c}\right) A(z) f_{\lambda,total}(\lambda_{eff,R})\right],
\label{eq:ch4_hgfluxlim}
\end{equation}
where $f_{\lambda,total}(\lambda_{eff,R})$ is the observed flux density per unit wavelength at $\lambda_{eff,R}$.  $A(z)$ is an aperture correction, to correct for the amount of host galaxy flux that does not fall within each SDSS optical fiber (which are 3$''$ in diameter.)  The fraction of light lost due to the finite fiber size can be significant, especially at lower redshifts where the host galaxy appears more extended.    We estimate the size of the galaxy as a function of redshift, assuming a de Vaucouleurs profile with effective radius $r_e=10$~kpc \citep[typical of \bl\ host galaxies, see][]{urry00_hstii}.  $A(z)$ is then approximately the total galaxy flux integrated over an infinite aperture divided by the amount of light expected to fall within 3$''$ at each redshift (accounting for cosmological effects.)    Note, since the AGN is a point source, no aperture correction needs to be applied to its flux estimate. 

We convert $f_{\nu,hg}(z)$ to a limit on the $R$-band host galaxy apparent magnitude, $R_{hg}(z)$, using the flux density zero point of the AB magnitude scale (3631~Jy at $\lambda_{eff,R}$) .  The redshift for which $R_{hg}(z)$ best matches the standard candle relation in Equation~2 of \citet{sbarufatti05} is then taken as a lower limit to the objects' redshift, $z_{hg}$.  We refer to these redshift limits throughout the text as ``host galaxy'' redshift limits.   The standard candle relation is only calibrated out to $z\sim0.7$, so we assign lower limits of $z_{hg}=0.7$ for 15 objects whose host galaxy apparent magnitudes are always fainter than expected from the standard candle relation.   Our approach is similar to the technique outlined in \citet{sbarufatti06}, who derive redshift limits by constraining AGN to host galaxy flux ratios in featureless \bl\ spectra using the equivalent width detection threshold of their high-resolution spectra.

The redshift distribution for all \noptall\ objects (i.e., including all spectroscopic redshifts and ``host galaxy'' redshift limits) is shown as open histograms in panels~(a) and (c) of Figure~\ref{fig:ch4_redshifts}.  The distribution of spectroscopic redshifts (assuming tentative and lower limit redshifts are exact) is shown as the hatched red histogram in panel~(a).  Lower limit spectroscopic redshifts (\noptallZll\ objects) are shown as the solid red histogram in that panel, and the inset shows a blowup of $z_{spec}<0.7$.  The distribution of ``host galaxy'' redshift limits (\noptallNoz\ objects) is shown as the blue hatched histogram in panel~(c), also with an inset showing a blowup of $z_{hg}<0.7$.   The insets only go out to $z=0.7$ because that is the maximum calibrated redshift of the standard candle relation.  For that reason, we also only compare the ``host galaxy'' redshifts to objects with $z_{spec}<0.7$ in the following.  

\begin{figure*}
\centering
\includegraphics[scale=0.65]{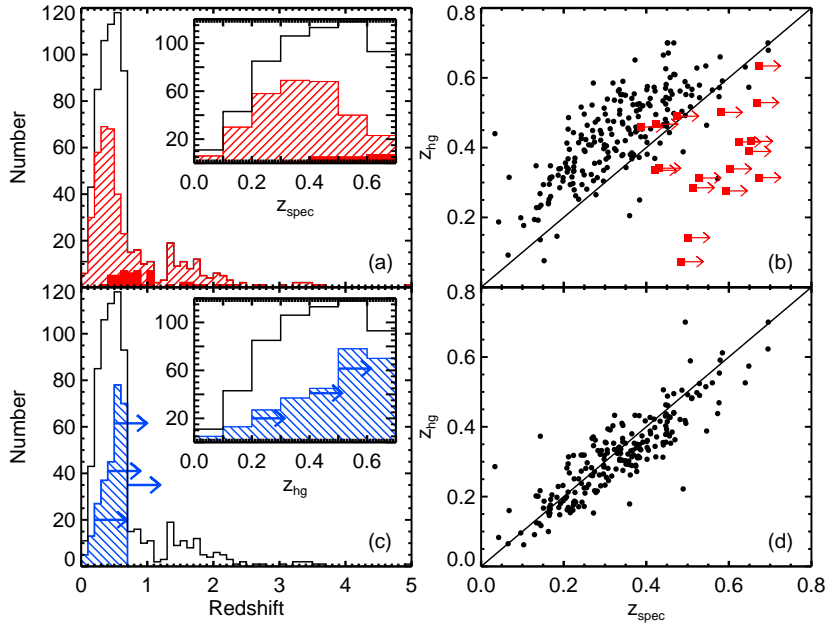}
\caption{Comparison between spectroscopic and host galaxy redshifts.   This figure is in color in the online version.  
(a) The redshift distribution of all \noptall\ \bl\ candidates (assuming redshift limits are exact).  The red hatched histogram shows the redshift distribution of the \noptallZtentPlusZrel\ objects with spectroscopic redshifts, and the red solid histogram shows the \noptallZll\ objects with lower limits from \ion{Mg}{2} absorption.  The inset shows a blowup of $0<z<0.7$.  
(b) {\it Black circles:} host galaxy redshift limits vs.\ spectroscopic redshifts for 213 objects with reliable $z_{spec}<0.7$.  Host galaxy flux limits are estimated with Equation~\ref{eq:ch4_hgfluxlim}, which underestimates most host galaxies' true fluxes.  Thus, $z_{hg}>z_{spec}$ for the vast majority.  {\it Red squares:} 18 objects with lower limit $z_{spec}<0.7$.  Host galaxies are not seen in their spectra, so the host galaxy redshift limits should be smaller than (or equal to) the spectroscopic redshift limits.  The line $z_{hg}=z_{spec}$ is drawn to guide the eye.  
(c) Same as panel (a), except the blue hatched histogram shows the $z_{hg}$ distribution for the \noptallNoz\ objects lacking spectroscopic redshifts.  Objects lacking spectroscopic redshifts tend to be higher-redshift on average.  
(d) Same as panel (b), except now the host galaxy flux estimates are exact (i.e., estimated with Equation~\ref{eq:ch4_hgflux}).  The spectroscopic and host galaxy redshifts agree rather well ($\sigma_{\Delta z}=\pm0.064$).
}
\label{fig:ch4_redshifts}
\end{figure*}

Even if one  assumes that  ``host galaxy'' redshifts are exact, the objects lacking spectroscopic redshifts are, on average, located at higher-redshift than objects with $z_{spec}<0.7$.  Considering the $\sim$200 objects with reliable $z_{spec}<0.7$, their average spectroscopic redshift is 0.33, compared to an average ``host galaxy'' redshift limit of 0.48 for the \noptallNoz\ objects lacking spectroscopic redshifts.  Later in this section we estimate that the ``host galaxy'' redshift limits are accurate to $\pm$0.064.  A Kolmogorov-Smirnov (K-S) test shows that the $z_{spec}$ distribution for objects with reliable $z_{spec}<0.7$ and the $z_{hg}$ distribution for objects lacking spectroscopic redshifts are statistically different; there is a negligible (10$^{-24}$) chance they are drawn from the same parent distribution.   There are likely some instances in our sample of extremely featureless spectra that are nearby; in this case the AGN must be highly beamed, and the finite size of the SDSS optical fiber limits the detectability of host galaxy spectral features.    However, the majority of extremely featureless spectra are likely among the most distant, which could have implications for understanding \bl\ properties (for example, their cosmic evolution) if not accounted for properly.

We check our ``host galaxy'' redshift algorithm by examining 18 \bl\ candidates with lower limit spectroscopic redshifts $z_{spec} < 0.7$ (i.e., spectroscopic redshift limits derived from \ion{Mg}{2} absorption).  We do not see obvious host galaxy features in their spectra, so we expect their ``host galaxy'' redshift limits to be smaller than (or approximately equal to) their spectroscopic redshift limits.  This is indeed true: the average difference between spectroscopic and ``host galaxy'' redshift limits is $\left<z_{spec}-z_{hg}\right>=0.18$ (with $z_{spec}-z_{hg}$ spanning $-0.072$ to 0.41).  These 18 objects are shown as red squares in Figure~\ref{fig:ch4_redshifts}b.  While arrows designating lower limits are only drawn for the spectroscopic redshifts, the ``host galaxy'' redshift limits for these 18 objects should also be considered lower limits.

We also test our algorithm using a subset of 213 \bl\ candidates with reliable spectroscopic  $z_{spec}<0.7$, and for which the host galaxy decomposition fit was preferred over just a power law.  Almost all of these objects  have host galaxy to nuclear flux ratios larger than 0.2 at 6581~\AA, so the above algorithm underestimates  their host galaxy flux densities and should therefore overestimate their redshifts.   These 213 objects are shown as black circles in Figure~\ref{fig:ch4_redshifts}b, which confirms that the majority ($>$90\%) of the ``host galaxy'' redshift limit estimates are larger than their reliable spectroscopic redshifts.   The line $z_{hg}=z_{spec}$ is drawn for guidance.  

Finally, a perhaps even better test is to estimate actual host galaxy flux densities for these 213 objects, and to then derive redshift estimates (instead of limits).  We estimate the host galaxy flux densities (per unit frequency) at 6581~\AA\ for these 213 objects as:
\begin{eqnarray}
 f_{\nu,hg}(z)  & = & \left(\frac{\lambda_{eff,R}^2}{c}\right) A(z) \times \nonumber \\
  & &  \left [f_{\lambda,total}(\lambda_{eff,R}) - f_{\lambda6165,0}(1.067)^{-\alpha_{\lambda}}\right], 
  \label{eq:ch4_hgflux}
\end{eqnarray}
\\ where the second term inside the brackets uses the power law fit parameters (referenced at 6165~\AA) returned by Equation~\ref{eq:ch4_decomp} to calculate the AGN's flux density at 6581~\AA\,.  We estimate the host galaxy flux density as the difference between the observed total flux density and the AGN flux density because it does not require {\it a priori} knowledge of the redshift.  We similarly do not use our knowledge of the spectroscopic redshift to estimate $A(z)$.  The spectroscopic and ``host galaxy'' redshifts match fairly well.  They are compared in Figure~\ref{fig:ch4_redshifts}d.  The line $z_{spec}=z_{hg}$ is overplotted for guidance (i.e., the solid line is not a fit to the data).  The ``host galaxy'' redshifts cluster along this line, with $\left<\Delta z\right>=\left<z_{spec}-z_{hg}\right>=0.015$ and a dispersion of $\sigma_{\Delta z}=\pm0.064$.   

In summary, we conclude from the above tests that our algorithm provides realistic redshift limits, and we estimate the ``host galaxy'' redshift limits to be accurate to $\pm0.064$ (i.e., $z>z_{hg}\pm0.064$).  We obtain similar uncertainties on the redshift limits if we add the uncertainties from our host galaxy flux measures to the $rms$ error of the standard candle relation.  We also note that the accuracy of our ``host galaxy'' redshift estimates is similar to uncertainties on ``imaging'' redshift limits placed by \citet{sbarufatti05} for objects lacking host galaxy detections with {\it HST}.

\subsubsection{AGN Component Optical Fluxes and Luminosities}
We also use the spectral fits to estimate optical fluxes and specific luminosities for the AGN component, excluding the host galaxy flux when contamination is significant.   For objects lacking spectroscopic redshifts, or for which the power law fit (Equation~\ref{eq:ch4_specind}) is preferred over the host galaxy decomposition (Equation~\ref{eq:ch4_decomp}), we assume the host galaxy does not contribute to the observed flux.   In these cases, we estimate each AGN optical flux as the psf magnitude in the filter whose effective wavelength\footnote{The SDSS filter effective wavelengths are 4686, 6165, 7481, and 8931~\AA\ for $g$,$r$,$i$, and $z$ respectively.} 
 is closest to 5000~\AA\ rest-frame.    These magnitudes are then corrected for Galactic extinction using the maps of \citet{schlegel98}. We assume the ``host galaxy'' redshift limit is exact to locate 5000~\AA\ rest-frame for objects lacking spectroscopic redshifts.  
 
 If the host galaxy decomposition is preferred over the power law fit, then that source also  has a spectroscopic redshift, which we use to identify the SDSS filter covering (or closest to) 5000~\AA\ rest-frame.  We use the AGN parameters from Equation~\ref{eq:ch4_decomp} to estimate the AGN component flux density at the chosen filter's effective wavelength.   We then convert the decomposed AGN flux densities to optical AB apparent magnitudes, and we correct for extinction.  The distribution of extinction corrected AGN component apparent magnitudes is shown in Figure~\ref{fig:ch4_fluxes}b.  Uncertainties on the decomposed AGN apparent magnitudes are typically around $0.03$~mag.
 
 Next, we use the above apparent magnitudes to estimate the AGN components' specific luminosities (per unit frequency) at 5000~\AA\ rest-frame (using each object's measured optical spectral index to perform the K-correction).  These luminosities are lower limits for objects with lower limit spectroscopic or ``host galaxy'' redshifts.
   
 \subsection{Radio Matches in FIRST and NVSS}
 \label{sec:ch4_radioinfo}
  We identify objects as radio emitters by correlating our \noptall\ \bl\ candidates to the FIRST and NVSS radio catalogs, and we find \noptallFirst\ matches within 2$''$ to a FIRST source and \noptallNvss\ matches within 10$''$ to an NVSS source.  Our choice of a 10$''$ search radius for NVSS is rather conservative, but we use it to avoid contamination from background radio sources (which increases rapidly for offsets $r>10''$.)   There are \noptallRadio\ objects that match to either a FIRST or NVSS radio source.  Among these \noptallRadio\ objects, \noptallFirstANDNvss\ objects match to {\it both} a FIRST and NVSS source, \noptallFirstNOTINNvss\ objects match only to a FIRST source, and \noptallNvssNOTINFirst\ match only to an NVSS source.
   
 We expect the number of random mismatches between FIRST-SDSS and NVSS-SDSS to be small.   Figure~\ref{fig:ch4_matchsep} shows cumulative distribution functions of  separations between optical and radio source positions, $\Delta r$, normalized to the relevant search radii, $r_{search}=2''$ and 10$''$ for FIRST and NVSS respectively.     The distribution functions are normalized to the total number of FIRST (\noptallFirst) and NVSS (\noptallNvss) radio matches.    We expect 90\% of FIRST-SDSS matches to have $\Delta r < 1''$ \citep{becker95}, and we estimate that $\sim$90\% of NVSS-SDSS matches (for NVSS sources with flux densities larger than 5~mJy) should have $\Delta r \lesssim 6''$ \citep[see][]{condon98}.  The above 90\% confidence radii (normalized to the relevant search radii) are overplotted as squares on each distribution function.  Both distribution functions show that $>90\%$ of FIRST-SDSS and NVSS-SDSS matches fall within $\Delta r=$ $1''$ and $6''$ respectively.  The horizontal dotted line in Figure~\ref{fig:ch4_matchsep} marks 90\% for guidance.  
 
 We further  investigate the chance of finding spurious matches to background radio sources by randomly offsetting the SDSS source positions by $\pm3-10$ times the relevant search radii for FIRST and NVSS.  We then recorrelate the (offsetted) \noptall\ \bl\ candidates to the FIRST and NVSS source catalogs.  We repeat this test 10 times, and on average we only find 2 (1) spurious matches to FIRST (NVSS) radio sources (i.e., we expect $<1\%$ of our sample to have spurious radio matches).   Thus, our radio identifications appear to be statistically reliable.
 
  \begin{figure}
\centering
\includegraphics[scale=0.5]{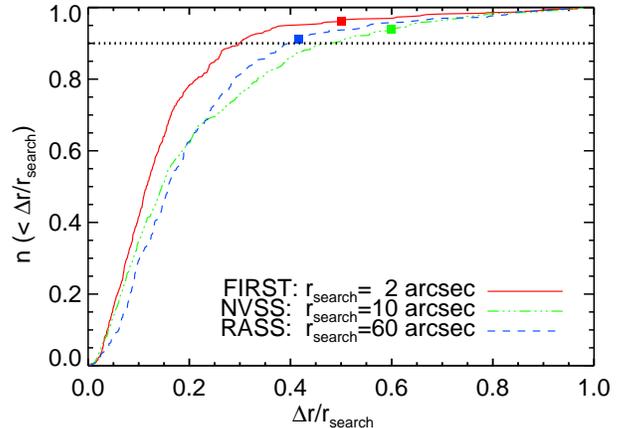}
\caption{Cumulative distribution functions of the fraction of objects with projected source separations, $\Delta r$, between FIRST-SDSS (solid red line), NVSS-SDSS (green dashed-dotted line) and RASS-SDSS (blue dashed line) source positions.  This figure is in color in the online version.  The abscissa is normalized to the relevant search radii used for each survey, (2$''$, 10$''$, and 60$''$ for FIRST, NVSS, and RASS respectively); the ordinate is normalized to the number of sources with matches in  each survey (\noptallFirst, \noptallNvss, and \noptallXray\ for FIRST, NVSS, and RASS respectively).  The solid squares mark the separations we expect to include $\sim$90\% of matches from independent studies of source positional offset distributions. These distribution functions indicate that our radio and X-ray identifications are statistically reliable.
}
\label{fig:ch4_matchsep}
\end{figure}

 \subsubsection{Radio Fluxes and Luminosities}
 We take the radio flux density at 1.4~GHz to be the integrated flux density given in the FIRST catalog for all FIRST matches (\noptallFirst\ objects).  Because of FIRST's better angular resolution, we use the FIRST flux density even if the object is detected in NVSS.   Since \bl\ objects are compact, FIRST's lower sensitivity to extended emission than NVSS (due to fewer short baselines) should not be important.   If an object only matches to an NVSS source (\noptallNvssNOTINFirst\ objects), then we use its NVSS integrated flux density.  If an object is not detected in either FIRST or NVSS but it is covered by the FIRST footprint (\noptallRadioLimits\ objects), then we place radio flux density limits from FIRST as $f_{1.4GHz}<0.25 + (5\sigma_{rms})$~mJy, where $\sigma_{rms}$ is the noise in the FIRST survey at the object's sky coordinates (typically $\sim$0.15~mJy~beam$^{-1}$).   The 0.25~mJy term is a correction for the CLEAN bias in the FIRST data reductions \citep[see][]{becker95}.  We do not place radio limits using NVSS (even for \noptallNoRadioInfo\ objects outside the FIRST footprint) due to the increased potential for source confusion.  The distribution of 1.4~GHz  radio flux densities (and limits) is shown in Figure~\ref{fig:ch4_fluxes}c.
  
 We then calculate radio specific luminosities (per unit frequency) at rest-frame 5~GHz, assuming \bl\ spectra follow a power law with index $\alpha_{\nu}=-0.27$ ($f_{\nu} \sim \nu^{-\alpha_{\nu}}$) in the radio, which is the average radio spectral index of the 1~Jy \bl\ radio sample \citep{stickel91}.  We again assume tentative and lower limit spectroscopic redshifts and ``host galaxy'' redshift limits are exact.

 \subsection{X-ray Matches in RASS}
 \label{sec:ch4_xrayinfo}
 We identify \noptallXray\ \bl\ candidates as X-ray emitters that match within 60$''$ to a RASS source (in the RASS Bright and Faint Source catalogs, \citealt{voges99,voges00}).  Given typical \bl\ SEDs and the flux sensitivities of RASS, SDSS, and FIRST, we expect most  \bl\ candidates detected in X-rays by RASS to also be detected in the radio (see Figure~1 of \citealt{plotkin08}). All but \noptallXrayNoRadio\ of the RASS detected \bl\ candidates are also detected in the radio by either FIRST or NVSS.  These 9 objects could be examples of extreme HBLs,  they could have happened to be in low flux states during the FIRST and NVSS observations (or in high flux states during the RASS observations), or some of these 9 objects might be viable radio-quiet \bl\ candidates, as even extant radio limits place about half of them in the radio-quiet regime (see \S \ref{sec:ch4_rquiet}).
  
 The cumulative distribution function of the RASS identifications is shown as the blue dashed line in  Figure~\ref{fig:ch4_matchsep}.  We expect 90\% of SDSS/RASS source matches to have $\Delta r < 25''$ \citep{voges99}; this is indeed observed, as indicated by the blue square marking $\Delta r=25''$ in Figure~\ref{fig:ch4_matchsep}.   We also test our X-ray identifications by randomly offsetting the SDSS source positions by $\pm 3-10'$ and recorrelating the (offsetted) \bl\ candidates to the RASS source catalog.  After 10 iterations, we find on average only 2 spurious matches (i.e., $<1\%$ of our sample), further indicating that  our \noptallXray\ X-ray identifications appear to be statistically reliable. 
 
 \subsubsection{X-ray Fluxes and Luminosities}
 We estimate X-ray fluxes and specific luminosities (or upper limits) from the RASS count rates.    We assume \bl\ objects have X-ray spectral indices of $\alpha_{\nu}=1.25$ \citep[a value approximately intermediate between LBLs and HBLs, see][]{sambruna96, sambruna97}, where the X-ray spectral index per unit frequency is again defined as $f_{\nu}\sim\nu^{-\alpha_{\nu}}$.  For the \noptallNoXray\ objects lacking X-ray detections, we take the exposure time for the closest match in the RASS X-ray catalog.  We then estimate upper limits for those objects' X-ray count rates as 6 photons divided by the adopted exposure time.\footnote{6 counts is the limit for inclusion in the RASS faint source catalog \citep[see][]{voges00}.}  
 We use the \citet{stark92} hydrogen maps and the {\tt colden} tool in CIAO \citep{ciaoref} to estimate the hydrogen column density along each object's sightline.  We then use PIMMS to convert ROSAT PSPC X-ray count rates (or count rate limits) to unabsorbed broad-band (0.1-2.4 keV) fluxes (or limits).  The distribution of broad-band X-ray fluxes and flux limits is shown in Figure~\ref{fig:ch4_fluxes}d.  We then convert the broad-band X-ray fluxes to X-ray specific luminosities (per unit frequency) at rest-frame 1~keV, again assuming tentative and lower limit spectroscopic redshifts and ``host galaxy'' redshift limits are exact.

One X-ray undetected source in RASS (SDSS J232428.43+144324.3, $z=1.41$) is weakly detected in our recent {\it Chandra} X-ray observation with 0.0012~counts~s$^{-1}$ from 0.5-6.0~keV.\footnote{The {\it Chandra} observation (\dataset[ADS/Sa.CXO#Obs/10386]{ObsId 10386}) was taken on 2009 May 31 with the Advanced CCD Imaging Spectrometer \citep[ACIS;][]{garmire03} at the nominal S3 aimpoint.}  
 We use its {\it Chandra} count rate to estimate it would have a RASS count rate of 0.0005~counts~s$^{-1}$ and an unabsorbed flux $1.25\times10^{-14}$~erg~s$^{-1}$~cm$^{-2}$ from 0.1-2.4~keV.  We assume a power law spectrum with photon index $\Gamma=2.25$ and a column density $N_H=4.15\times10^{20}$~cm$^{-2}$ \citep{stark92}.  The inferred RASS count rate is well below the limit for inclusion in the RASS catalog.  This source is outside of the FIRST footprint, and it is undetected by NVSS in the radio.  It also remained undetected at 8.4~GHz in follow-on observations with the Very Large Array (VLA), with radio flux density $f_{8.4~GHz} < 0.3$~mJy.\footnote{This source was observed in the VLA D configuration on 2007 March 30 (Program ID AP524).  The VLA is operated by the National Radio Astronomy Observatory, a facility of the National Science Foundation operated under cooperative agreement by Associated Universities, Inc.} 

\subsection{Broad-band Spectral Indices}
\label{sec:ch4_alphas}
Finally, we use the specific luminosities at each waveband to calculate the broad-band spectral indices $\alpha_{ro}$ and $\alpha_{ox}$ (or limits in the cases of radio/X-ray non-detections).\footnote{The broad-band spectral index, for $\nu_2>\nu_1$, is defined as $\alpha_{\nu_1\nu_2}=-\log(L_{\nu_2}/L_{\nu_1})/\log(\nu_2/\nu_1)$.  Here, $\alpha_{ro}=-\log(L_o/L_r)/5.08$, $\alpha_{ox}=-\log(L_x/L_o)/2.60$, and $\alpha_{rx}=-\log(L_x/L_r)/7.68$, where $L_r$, $L_o$, and $L_x$~are the specific luminosities (per unit frequency) at 5 GHz, 5000 \AA, and 1 keV respectively \citep{tananbaum79, stocke85}.}  
  We note that these values are relatively insensitive to the exact redshift used for their calculation.   The $\alpha_{ro}-\alpha_{ox}$ plane is shown in Figure~\ref{fig:ch4_alpha}.  Objects with RASS X-ray detections are shown as blue triangles, and objects lacking X-ray detections are shown as black circles.  Red arrows designate $\alpha_{ro}$ upper limits for objects lacking radio detections in FIRST/NVSS.  All black circles are $\alpha_{ox}$ lower limits, but for ease of display we do not draw arrows for those limits in Figure~\ref{fig:ch4_alpha}.   The blue square marks SDSS J2324+1443 ($\alpha_{ro}<0.003$, $\alpha_{ox}=1.796$) using its deeper {\it Chandra} X-ray and VLA radio observations.  The dashed line marks the boundary between Low- and High-Energy Peaked \bl\ objects \citep[LBLs and HBLs, respectively, see][]{padovani95_apj}.\footnote{HBLs are commonly defined as $\alpha_{rx}\le0.75$ corresponding to $F_x/F_r\gtrsim10^{-11}$~with the X-ray flux in erg s$^{-1}$~cm$^{-2}$~and the radio flux in Jy \citep{padovani95_apj}.  The dividing line is also sometimes quoted as the similar $F_x/F_r\ge10^{-5.5}$~where both fluxes are in Jy \citep{wurtz94,perlman96}.}

 \begin{figure*}
\centering
\includegraphics[scale=0.65]{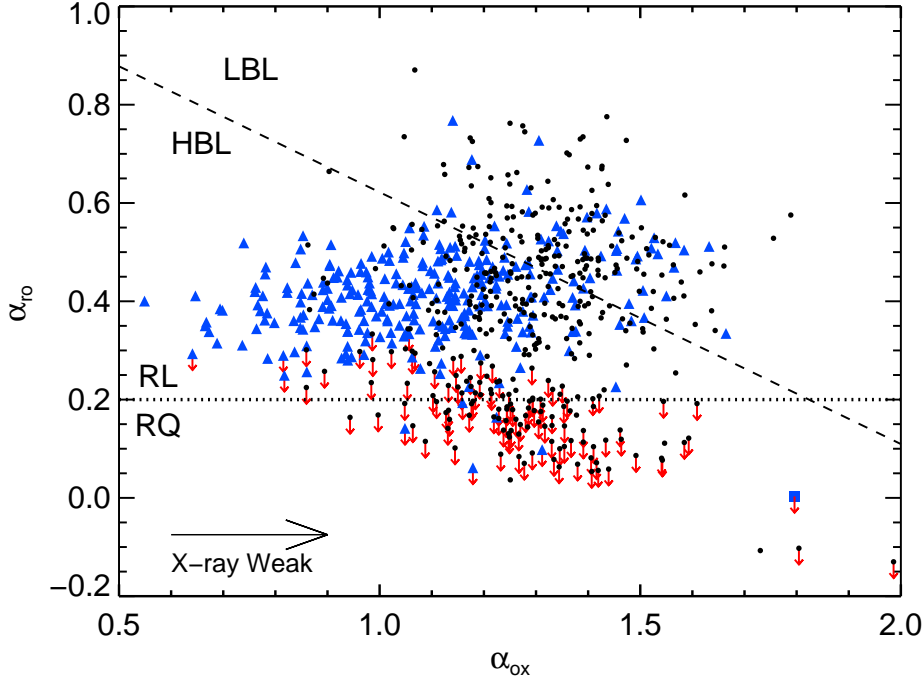}
\caption{Broad-band spectral indices $\alpha_{ro}$ vs.\ $\alpha_{ox}$.  This figure is in color in the online version.  The dotted line shows our adopted division between radio-loud (RL) and radio-quiet (RQ) AGN, and the dashed line shows the typical division ($\alpha_{rx}=0.75$) between HBLs and LBLs.   Red arrows show upper limits on $\alpha_{ro}$ for objects lacking radio detections in FIRST/NVSS.  
{\it Blue triangles:} Objects that match to an X-ray source in RASS. 
{\it Blue square:} Object with a weak X-ray detection in a follow-on {\it Chandra} observation. 
{\it Black circles:} Objects that do not match to a RASS X-ray source.  All black circles are actually lower limits on $\alpha_{ox}$, but to avoid cluttering the figure we do not illustrate these limits with arrows.   
There is a continuous population of \bl\ objects, with a smooth transition between HBLs and LBLs.  However, the SDSS might be biased toward recovering excess HBLs.  We also recover a large population of radio-quiet objects with featureless spectra; this population probably consists of a mix of  \bl\ objects lying on the radio-weak tail of the parent distribution and another class of featureless objects like WLQs.   Most of the recovered radio-quiet objects have $z<2.2$, so they would constitute the largest sample to date of such exotic objects at these redshifts.  Some of the radio-quiet objects may populate a similar region of the $\alpha_{ro}-\alpha_{ox}$ plane as high-redshift WLQs \citep{shemmer09}.
}
\label{fig:ch4_alpha}
\end{figure*}

The SDSS covers a large enough area to recover rare populations of objects, such as radio-weak AGN with featureless spectra.   The boundary between radio-loud and radio-quiet quasars is usually taken as a radio loudness parameter (i.e., radio to optical flux ratio) of $R=10$ \citep[see][]{kellermann89, stocke92}.  This approximately corresponds to $\alpha_{ro}=0.2$ (marked with a dotted line in Figure~\ref{fig:ch4_alpha}), and we recover \noptrq\ featureless spectra with $\alpha_{ro}<0.2$ (not including the \noptallNoRadioInfo\ objects outside the FIRST footprint.)   We identify objects as radio-loud or radio-quiet assuming upper $\alpha_{ro}$ limits are exact.  Only \noptRquietRadioDetection\ of the \noptrq\ radio-quiet objects have radio-detections, so the radio-weakness of many objects is even more extreme than we assume in this analysis.  Sample spectra of 4 of these radio-quiet objects are shown in Figure~\ref{fig:ch4_rqspec}.

\begin{figure*}
\centering
\includegraphics[scale=0.65]{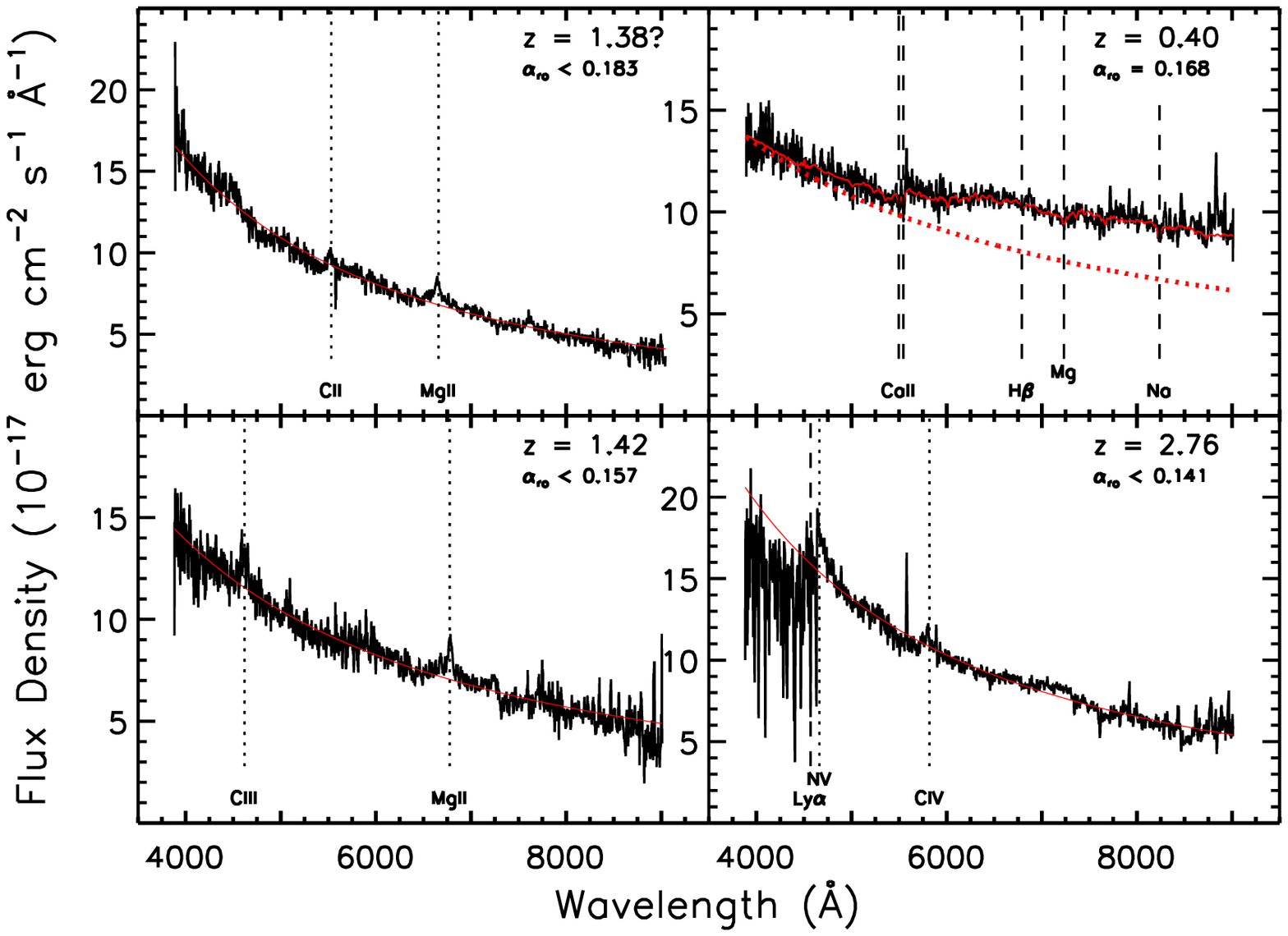}
\caption{Sample spectra of 4 weak-featured radio-quiet objects with selected spectral features labeled.  Absorption features are marked with dashed lines, and emission features are marked with dotted lines.   No object matches to a RASS X-ray source, and values for $\alpha_{ro}$ are given in each panel.   The spectral fits from the host galaxy decomposition described in \S \ref{sec:ch4_specdecomp} are overplotted as red lines.  If there is significant host galaxy contamination, then the dotted red lines illustrate just the power law flux from the AGN.  
{\it Top left:} SDSS J103806.71$+$382626.5.  This object has a tentative spectroscopic redshift assuming the strongest emission feature is \ion{Mg}{2}.  Although we assign this a tentative spectroscopic redshift, weak \ion{C}{2}] may have also been detected. 
{\it Top right:} SDSS J110938.50$+$373611.6.  This object has a reliable spectroscopic redshift and is detected in the radio by FIRST (with $\alpha_{ro}=0.168$ placing it in the radio-quiet regime.)  This may be an example of a \bl\ object lying on the radio-weak tail of the parent distribution. 
{\it Bottom left:} SDSS J132138.86$+$010846.2, an object with a reliable spectroscopic redshift.   
{\it Bottom right:} SDSS J233939.48$-$103539.3, a high-redshift object that shows the Ly$\alpha$ forest, and weak \ion{N}{5} and \ion{C}{4} emission.  This object is typical of other SDSS high-redshift WLQs, and it is additionally listed  in Table~\ref{tab:ch4_wlq} as a new WLQ identification.
}
\label{fig:ch4_rqspec}
\end{figure*}

Some of these radio-weak objects might be examples of beamed radio galaxies that simply populate the radio-faint tail of the much larger radio-loud SDSS \bl\ population.  However, it is also possible that the SDSS is large enough to recover another rare population of objects with featureless optical spectra.  In this case, the radio-bright tail of the alternate population would overlap with the radio-faint tail of the \bl\ population, making a secure \bl\ classification difficult with extant data.  The vast majority of these objects have spectroscopic redshifts, and their average optical luminosity of $\nu L_{\nu} \sim 10^{46}$~erg~s$^{-1}$ (at rest-frame 5000~\AA; all \noptrq\ objects have $\nu L_{\nu} >10^{43}$~erg~s$^{-1}$) indicates that these are almost definitely AGN.   We compare and contrast these odd objects with the larger radio-loud \bl\ sample in \S \ref{sec:ch4_rquiet}.    

These \noptrq\ objects constitute an interesting population, whether or not they are not best described as \bl\ objects.  For example, objects lying on the high-redshift tail ($z>2.2$) might be classified as  WLQs.  Their emission lines are plausibly intrinsically weak or absent rather than being washed out by strong relativistically boosted jets \citep{shemmer06,shemmer09,diamond09}.  However, it is possible that some of these objects still possess weak jets.  Also, SDSS J2324+1443 (the blue square in Figure~\ref{fig:ch4_alpha}) suggests that deeper X-ray observations  of  some of these objects may show them to have dissimilar X-ray properties compared to the radio-loud \bl\ candidates.  We thus separate objects with $\alpha_{ro}<0.2$ from the rest of the \bl\ candidates, and we refer to this population as weak-featured radio-quiet objects.   

\subsection{Catalogs of Optically Selected \bl\ Candidates}
\label{optcat}
The final optically selected sample is presented in Tables~\ref{tab:ch4_empdata_rl} -- \ref{tab:ch4_deriveddata_rq}.  Only the initial 5 entries of each table are shown here; the full versions are available in the electronic version of this article.  Table~\ref{tab:ch4_empdata_rl} includes primarily empirical data from the SDSS database for the \noptrl\ radio-loud \bl\ candidates.  The IAU designation, right ascension, and declination of each source are given in the first 3 columns.  Our classification as a high-confidence (`H') or low-confidence (`L') \bl\ candidate is given in column (4).  Column (5) indicates whether each source is identified as a \bl\ object in NED (as of May 2009): `Y' means it is, and `N' indicates it is not.   Columns (6)-(10) give SDSS psf $u,g,r,i,z$ band magnitudes.   The spectroscopic redshift ($z_{spec}$) is given in column (11), and column (12) indicates if the redshift is reliable (`R'), tentative (`T'), a lower limit from a \ion{Mg}{2} absorption doublet (`L'), or unknown (`U').  The measured \ion{Ca}{2}~H/K depression, $C$,  is given in column (13).  Column (14) gives the AGN optical spectral index $\alpha_{\nu}$ measured in \S \ref{sec:ch4_specdecomp}.  Finally, column (15) notes whether a source is published as an SDSS \bl\ candidate in \citet[][`C05']{collinge05}, \citet[][`A07']{anderson07} and/or \citet[][`P08']{plotkin08}.  This information is listed because not all SDSS \bl\ candidates are identified as such in NED.    We also note objects identified as $z>2.7$ WLQs in \citet[][`S09']{shemmer09} or in \citet[][`D09']{diamond09}.  Table~\ref{tab:ch4_empdata_rq} presents the same information for the \noptrq\ weak-featured radio-quiet objects.  

\tabletypesize{\tiny}
\begin{deluxetable*}{lrrccccccccccrl}
\tablewidth{0pt}
\tablecaption{Observed Parameters of Optically Selected Radio-loud BL Lac Candidates\label{tab:ch4_empdata_rl}}
\tablehead{
           \colhead{SDSS Name} 
        & \colhead{RA}     
        & \colhead{Dec}
        &  
        &  \colhead{in} 
        & 
        & 
        & 
         & 
         & 
         & 
         & \colhead{Redshift}
         & 
         & 
         &  
\\
           \colhead{(J2000)}  
        & \colhead{(J2000)}  
        & \colhead{(J2000)} 
        & \colhead{Classif.}   
        & \colhead{NED?} 
        & \colhead{$u$} 
        & \colhead{$g$} 
        & \colhead{$r$} 
        & \colhead{$i$} 
        & \colhead{$z$} 
        & \colhead{$z_{spec}$} 
        & \colhead{Flag} 
        & \colhead{$C$} 
        & \colhead{$\alpha_{\nu}$} 
        & Comments 
\\
         \colhead{(1)}
         & \colhead{(2)}
         & \colhead{(3)}
         & \colhead{(4)}
         & \colhead{(5)}
         & \colhead{(6)}
         & \colhead{(7)}
         & \colhead{(8)}
         & \colhead{(9)}
         & \colhead{(10)}
         & \colhead{(11)}
         & \colhead{(12)}
         & \colhead{(13)}
         & \colhead{(14)}
         & (15)
}
\startdata
000157.23$-$103117.3  &   0.48850  & -10.52148  &    H  &     N  & 20.66  & 19.74  & 18.77  & 18.26  & 17.86  &   0.252  & R  &   0.393  &  1.62  &               P08 \\
000257.17$-$002447.3  &   0.73824  &  -0.41315  &    H  &     N  & 21.10  & 20.63  & 19.89  & 19.22  & 18.85  &   0.522  & R  &   0.247  &  2.35  &               P08 \\
001736.90$+$145101.9  &   4.40378  &  14.85053  &    H  &     Y  & 18.80  & 18.33  & 17.86  & 17.56  & 17.28  &   0.303  & R  &   0.132  &  1.16  &               C05 \\
002142.25$-$090044.4  &   5.42608  &  -9.01234  &    H  &     N  & 20.15  & 19.79  & 19.32  & 18.82  & 18.52  &   0.648  & R  &   0.218  &  1.30  &           C05;P08 \\
002200.95$+$000657.9  &   5.50396  &   0.11610  &    H  &     Y  & 20.48  & 20.04  & 19.28  & 18.87  & 18.55  &   0.306  & R  &   0.356  &  1.38  &       C05;A07;P08 \\
\enddata
\tablecomments{Table~\ref{tab:ch4_empdata_rl} is available in its entirity in the electronic version of this article.  A portion is shown here for guidance regarding its form and content.}
\end{deluxetable*}
 
\tabletypesize{\tiny}
\begin{deluxetable*}{lrrccccccccccrl}
\tablewidth{0pt}
\tablecaption{Observed Parameters of Weak-featured Radio-quiet Objects\label{tab:ch4_empdata_rq}}
\tablehead{
           \colhead{SDSS Name} 
        & \colhead{RA}     
        & \colhead{Dec}
        &  
        &  \colhead{in} 
        & 
        & 
        & 
         & 
         & 
         & 
         & \colhead{Redshift}
         & 
         & 
         &  
\\
           \colhead{(J2000)}  
        & \colhead{(J2000)}  
        & \colhead{(J2000)} 
        & \colhead{Classif.}   
        & \colhead{NED?} 
        & \colhead{$u$} 
        & \colhead{$g$} 
        & \colhead{$r$} 
        & \colhead{$i$} 
        & \colhead{$z$} 
        & \colhead{$z_{spec}$} 
        & \colhead{Flag} 
        & \colhead{$C$} 
        & \colhead{$\alpha_{\nu}$} 
        & Comments 
\\
         \colhead{(1)}
         & \colhead{(2)}
         & \colhead{(3)}
         & \colhead{(4)}
         & \colhead{(5)}
         & \colhead{(6)}
         & \colhead{(7)}
         & \colhead{(8)}
         & \colhead{(9)}
         & \colhead{(10)}
         & \colhead{(11)}
         & \colhead{(12)}
         & \colhead{(13)}
         & \colhead{(14)}
         & (15)
}
\startdata
001741.84$-$105613.2  &   4.42437  & -10.93703  &    L  &     N  & 19.21  & 18.98  & 18.79  & 18.64  & 18.64  &   1.806  & R  & \nodata  &  0.62  &     \nodata             \\
005713.01$+$004205.9  &  14.30421  &   0.70164  &    H  &     N  & 19.94  & 19.48  & 19.11  & 18.89  & 18.80  &   1.538  & T  & \nodata  &  1.24  &      \nodata             \\
025716.22$-$001433.5  &  44.31761  &  -0.24265  &    L  &     N  & 19.96  & 19.73  & 19.46  & 19.25  & 19.19  &   1.919  & T  & \nodata  &  0.61  &      \nodata             \\
025743.73$+$011144.5  &  44.43222  &   1.19572  &    L  &     N  & 19.60  & 19.15  & 18.89  & 18.76  & 18.69  &   1.712  & T  & \nodata  &  1.01  &      \nodata             \\
075331.84$+$270415.3  & 118.38269  &  27.07094  &    L  &     N  & 19.18  & 18.90  & 18.66  & 18.57  & 18.59  &   1.655  & R  & \nodata  &  0.74  &     \nodata              \\
\enddata
\tablecomments{Table~\ref{tab:ch4_empdata_rq} is available  in its entirity in the electronic version of this article.  A portion is shown here for guidance regarding its form and content.}
\end{deluxetable*}

Table~\ref{tab:ch4_deriveddata_rl} includes primarily derived multiwavelength data for the \noptrl\ radio-loud \bl\ candidates.  The SDSS source name and spectroscopic redshifts ($z_{spec}$) are repeated in columns (1) and (2) for easy comparison to Table~\ref{tab:ch4_empdata_rl}.  Column (3) includes our ``host galaxy'' redshift limits for objects lacking spectroscopic redshifts.   An estimate of the extinction corrected magnitude of just the flux from the AGN component is given in Column (4) (in the SDSS filter closest to 5000~\AA\ rest-frame).   The logarithm of the corresponding AGN component specific luminosity at rest-frame 5000~\AA\ (in units of erg~s$^{-1}$~Hz$^{-1}$) is given in Column (5).  Column (6) includes a flag noting the radio survey from which we derive radio fluxes and luminosities.  `F' indicates the radio information comes from a detection in the FIRST survey; `N' indicates the radio information comes from a detection in the NVSS survey; `U' indicates that the source was undetected by both FIRST and NVSS, and our radio fluxes and luminosities are upper limits derived from FIRST; `X' means the object is outside of the FIRST footprint, and we do not include any radio information.  Column (7) gives integrated radio flux densities at 1.4~GHz in mJy, and column (8) gives the corresponding logarithm of their specific luminosities at rest-frame 5~GHz (in units erg~s$^{-1}$~Hz$^{-1}$).  The values in columns (7) and (8) are upper limits if the radio flag in column (6) is set to `U', and those entries are blank if the radio flag is set to `X'.  Column (9) contains a flag noting whether a source is detected (`R') or undetected (`U')  in the X-ray by RASS.  Column (10) lists the broad-band unabsorbed X-ray flux from 0.1-2.4~keV (in units of $10^{-13}$~erg~s$^{-1}$~cm$^{-2}$), and the logarithm of the specific X-ray luminosity at 1 keV (in units of erg~s$^{-1}$~Hz$^{-1}$) is given in column (11).    Columns (10) and (11) are upper limits if the X-ray flag in column (9) is set to `U'.  Finally, broad-band spectral indices between 5~GHz and 5000~\AA\ ($\alpha_{ro}$), and 5000~\AA\ and 1~keV ($\alpha_{ox}$) are given in columns (12) and (13) respectively.   If the radio flag in column (6) is set to `U', then $\alpha_{ro}$ is an upper limit, and $\alpha_{ox}$ is a lower limit if the X-ray flag in column (9) is set to `U'.   The same information is presented in Table~\ref{tab:ch4_deriveddata_rq} for the \noptrq\ weak-featured radio-quiet objects.

\tabletypesize{\tiny}
\begin{deluxetable*}{lcrccc r@{.}l  ccrccc}
\tablewidth{0pt}
\tablecaption{Derived Parameters of Optically Selected Radio-loud BL Lac Candidates\label{tab:ch4_deriveddata_rl}}
\tablehead{
	    \colhead{SDSS Name} 
	& 
	&  
   	& 
	&  
	& \colhead{Radio} 
	& \multicolumn{2}{c}{$f_{1.4\ GHz}$} 
	&   
  	& \colhead{X-ray} 
	& 
	 & 
	 &  
	 & 
\\			   
	    \colhead{(J2000)}  
	& \colhead{$z_{spec}$} 
	& \colhead{$z_{hg}$}        
	&  \colhead{$m_{0,AGN}$}   
	&  \colhead{$\log L_{5000~{\rm \AA}}$}   
	&  \colhead{Flag}  
	& \multicolumn{2}{c}{(mJy)} 
	&  \colhead{$\log L_{5\ GHz}$} 
	& \colhead{Flag}  
	& \colhead{$F_X$}  
	&  \colhead{$\log L_{1\ keV}$}   
	&  \colhead{$\alpha_{ro}$}  
	&  \colhead{$\alpha_{ox}$}  
\\
	    \colhead{(1)}    
	 & \colhead{(2)}      
	 &   \colhead{(3)} 
	 &   \colhead{(4)}  
	 &   \colhead{(5)}  
	 &   \colhead{(6)}  
	 &   \multicolumn{2}{c}{(7)} 
	 &   \colhead{(8)} 
	 &   \colhead{(9)}  
	 &   \colhead{(10)}  
	 &  \colhead{(11)} 
	 &  \colhead{(12)} 
	 &  \colhead{(13)}
}
\startdata
000157.23$-$103117.3  &   0.252  & \nodata  &   18.17  &  29.347  &  F  &    32&970  &    31.822  &  U  &     3.811  &    25.911  &   0.487  &   1.320   \\
000257.17$-$002447.3  &   0.522  & \nodata  &   18.80  &  29.722  &  F  &   159&130  &    33.146  &  U  &     3.373  &    26.627  &   0.674  &   1.188   \\
001736.90$+$145101.9  &   0.303  & \nodata  &   17.73  &  29.843  &  N  &    58&739  &    32.236  &  U  &     6.946  &    26.361  &   0.471  &   1.337   \\
002142.25$-$090044.4  &   0.648  & \nodata  &   18.46  &  30.168  &  F  &    41&960  &    32.752  &  U  &     3.976  &    26.936  &   0.509  &   1.241   \\
002200.95$+$000657.9  &   0.306  & \nodata  &   19.21  &  29.268  &  F  &     1&730  &    30.714  &  R  &    21.140  &    26.854  &   0.285  &   0.927   \\
\enddata
\tablecomments{Table \ref{tab:ch4_deriveddata_rl} is available in its entirety in the electronic version of this article.  A portion is shown here for guidance regarding its form and content.}
\end{deluxetable*}

\tabletypesize{\tiny}
\begin{deluxetable*}{lcrccccccrccc}
\tablewidth{0pt}
\tablecaption{Derived Parameters of Weak-featured Radio-quiet Objects\label{tab:ch4_deriveddata_rq}}
\tablehead{
	    \colhead{SDSS Name} 
	& 
	&  
   	& 
	&  
	& \colhead{Radio} 
	& \colhead{$f_{1.4\ GHz}$} 
	&   
  	& \colhead{X-ray} 
	& 
	 & 
	 &  
	 & 
\\			   
	    \colhead{(J2000)}  
	& \colhead{$z_{spec}$} 
	& \colhead{$z_{hg}$}        
	&  \colhead{$m_{0,AGN}$}   
	&  \colhead{$\log L_{5000~{\rm \AA}}$}   
	&  \colhead{Flag}  
	& \colhead{(mJy)} 
	&  \colhead{$\log L_{5\ GHz}$} 
	& \colhead{Flag}  
	& \colhead{$F_X$}  
	&  \colhead{$\log L_{1\ keV}$}   
	&  \colhead{$\alpha_{ro}$}  
	&  \colhead{$\alpha_{ox}$}  
\\
	    \colhead{(1)}    
	 & \colhead{(2)}      
	 &   \colhead{(3)} 
	 &   \colhead{(4)}  
	 &   \colhead{(5)}  
	 &   \colhead{(6)}  
	 &   \colhead{(7)} 
	 &   \colhead{(8)} 
	 &   \colhead{(9)}  
	 &   \colhead{(10)}  
	 &  \colhead{(11)} 
	 &  \colhead{(12)} 
	 &  \colhead{(13)}
}
\startdata
001741.84$-$105613.2  &   1.806  & \nodata  &   18.68  &  31.222  &  U  &     1.785  &    32.195  &  U  &     4.537  &    28.158  &   0.192  &   1.176   \\
005713.01$+$004205.9  &   1.538  & \nodata  &   19.02  &  31.123  &  U  &     0.915  &    31.787  &  U  &     3.608  &    27.874  &   0.131  &   1.247   \\
025716.22$-$001433.5  &   1.919  & \nodata  &   19.27  &  31.047  &  U  &     0.780  &    31.879  &  U  &    10.450  &    28.590  &   0.164  &   0.943   \\
025743.73$+$011144.5  &   1.712  & \nodata  &   18.65  &  31.318  &  U  &     0.995  &    31.902  &  U  &    11.100  &    28.486  &   0.115  &   1.088   \\
075331.84$+$270415.3  &   1.655  & \nodata  &   18.57  &  31.224  &  U  &     0.940  &    31.853  &  U  &     3.711  &    27.971  &   0.124  &   1.249   \\
\enddata
\tablecomments{Table \ref{tab:ch4_deriveddata_rq} is available in its entirety in the electronic version of this table.  A portion is shown here for guidance regarding its form and content.}
\end{deluxetable*}

There is some overlap between the phenomenological properties of \bl\ objects and WLQs.  For example, 6 of the serendipitous $z>2.2$ WLQ discoveries listed in Table~\ref{tab:ch4_wlq} additionally pass our \bl\ criteria (i.e., their emission features show measured $REW<5$~\AA).  Those 6 objects are also included as weak-featured radio-quiet objects in Tables~\ref{tab:ch4_empdata_rq} and \ref{tab:ch4_deriveddata_rq}. 

\section{Discussion}
\label{sec:ch4_discussion}
\subsection{The Radio-Loud Subset}
\label{sec:ch4_rlsubset}
The properties of the radio-loud subset are consistent with the results presented in \citet{plotkin08}.  We refer the reader to that paper for a more detailed discussion: there is a smooth transition between LBLs and HBLs, and we also see a potential SDSS bias toward recovering HBLs.   However, we note that many of the RASS X-ray flux limits are not sensitive enough to confirm this bias.  Also, the radio-loud \bl\ candidates tend to appear more luminous as the strength of the \ion{Ca}{2}~H/K break decreases (i.e., as one views the relativistic jet more directly along its axis, also see \citealt{landt02}).   These properties are interpreted as generally  supporting the standard orientation based unification paradigm. 
 
\subsection{What are the Weak-Featured Radio-Quiet Objects?}
\label{sec:ch4_rquiet}
The population of \noptrq\ radio-quiet objects with nearly featureless spectra is intriguing.  Here we discuss the possibility that these are a distinct population from the \noptrl\ radio-loud \bl\ candidates.  These objects are at systematically higher redshift than their radio-loud counterparts.  This is illustrated in Figure~\ref{fig:ch4_rqdist}a, which shows the redshift distribution of all \noptall\ \bl\ candidates.  The dotted lines mark a region that is blown up in the inset, which shows the redshifts of the \noptrq\ radio-quiet objects color-coded by their redshift.  The blue histogram is for the \noptRquietLowZ\ objects with $z<1$.  ``Host galaxy'' redshifts are shown as the black filled histogram for \noptrqNoz\ radio-quiet objects lacking spectroscopic redshifts (all other redshifts for the radio-quiet objects are derived spectroscopically.)  Mid-range redshifts ($1<z<2.2$) are shaded green, and large redshifts ($z>2.2$) are shaded red.  The open histogram in the inset shows the distribution of the entire sample for reference.  

 \begin{figure*}
\centering
\includegraphics[scale=0.65]{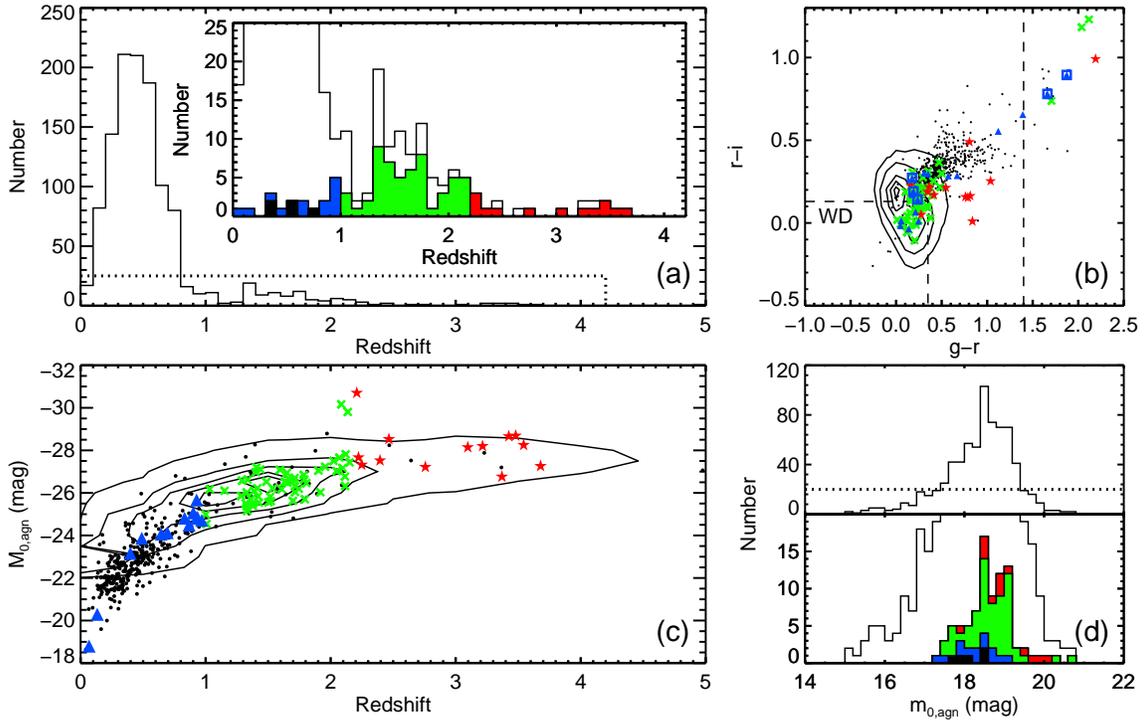}
\caption{  
(a) The redshift distribution of all \noptall\ nearly featureless spectra.  The inset shows a blowup (marked by dotted lines in the larger panel) with $z<1$, $1<z<2.2$, and $z>2.2$ radio-quiet objects colored blue, green, and red respectively.  The black filled histogram shows ``host galaxy'' redshifts for \noptrqNoz\ radio-quiet objects lacking spectroscopic redshifts.  The radio-quiet objects are at systematically higher redshifts.  
(b) $r-i$ vs.\ $g-r$, not corrected for extinction or host galaxy contamination; post-visual inspection color-cuts are overdrawn with dashed lines (see \S \ref{sec:ch4_postvisinspec}).    {\it Black circles}: radio-loud \bl\ candidates; {\it blue triangles, green crosses}, and {\it red star symbols}: weak-featured radio-quiet objects with spectroscopic redshifts $z<1$, $1<z<2.2$, and $z>2.2$, respectively.  The optical colors of $\sim$38,000 SDSS quasars with $1<z<2.2$ that lack radio detections in FIRST are shown as contours (contours mark 100, 500, 1500, 2500, and 3500 quasars).   The weak-featured radio-quiet objects cover the full range of optical colors as the radio-loud objects, and the mid-redshift weak-featured radio-quiet objects in general have similar optical colors as normal $1<z<2.2$ quasars.  None of the radio-quiet objects lacking spectroscopic redshifts (shown as blue triangles circumscribed by squares) fall in the $gri$ box aimed at removing DC white dwarfs (labeled WD).    
(c) Absolute magnitude vs.\ spectroscopic redshift.  AGN component absolute magnitudes in the filter closest to 5000~\AA\ rest-frame are shown for the radio-loud and radio-quiet objects with spectroscopic redshifts (symbols are the same as panel b).  Contours show $\sim$65,000 spectroscopically confirmed SDSS quasars without a match to a FIRST radio source \citep{schneider07}; contours mark 100, 1000, 2000, 3000, and 4000 quasars.  Absolute magnitudes are taken from \citet[][i.e., in the $i$ filter, assuming an optical spectral index per unit frequency of 0.5, and adopting the same cosmology as this study]{schneider07}.   The luminosities of the nearly featureless spectra are similar to normal quasars, indicating significant overlap in their optical properties (other than their weak or absent emission features).
(d) Distribution of decomposed AGN optical apparent magnitudes.  The top panel shows all \noptall\ objects, and the bottom panel is a blowup (marked by the dotted line in the top panel).  The distributions of radio-quiet objects at different redshifts are offset for clarity, and they follow the same coloring scheme as panel~(a).
}
\label{fig:ch4_rqdist}
\end{figure*}

Only \noptrqNoz\ radio-quiet objects lack spectroscopic redshifts, so there is not much concern for stellar contamination to the \noptrq\ object $\alpha_{ro}<0.2$ subset.  Plus, those \noptrqNoz\ objects lie outside of the $gri$ box designed to remove stellar contaminants (see \S \ref{sec:ch4_dcwd}), as is shown in Figure~\ref{fig:ch4_rqdist}b.  Plotted are $r-i$ vs.\ $g-r$ colors, not corrected for extinction or host-galaxy contamination; the color scheme is the same as in panel~(a), with the \noptrl\ object radio-loud subset shown as black circles for comparison.  The radio-quiet objects cover the same range of optical colors as the radio-loud \bl\ objects, further indicating significant overlap, at least phenomenologically, between the radio-loud and radio-quiet objects even if they are indeed two separate populations.  It does appear, however, that the optical colors of many of the $1<z<2.2$ radio-quiet objects tend to populate the bluer end of the radio-loud \bl\ distribution.   But, their colors are typical of normal spectroscopically confirmed SDSS quasars \citep{schneider07} at similar redshifts that lack radio detections in FIRST (see contours in Figure~\ref{fig:ch4_rqdist}b).  

We further explore the optical properties of the weak-featured radio-quiet objects by examining their optical luminosities as a function of redshift (Figure~\ref{fig:ch4_rqdist}c, which follows the same color scheme as panel~b).  The optical luminosities of the weak-featured radio-quiet objects follow the same trend with redshift as the radio-loud \bl\ objects.  Furthermore, the radio-quiet objects' luminosities are similar to the optical luminosities of $\sim$65,000 normal spectroscopically confirmed SDSS quasars that lack radio detections in FIRST.  We thus conclude that, other than their lack of strong emission features, the weak-featured radio-quiet objects have similar optical properties to both normal quasars and to radio-loud \bl\ objects. 

The radio-loud and radio-quiet objects also follow similar optical flux distributions, as illustrated in Figure~\ref{fig:ch4_rqdist}d.  The top panel shows the distribution of decomposed (extinction corrected) AGN optical magnitudes for the entire \noptall\ object sample, and a zoom in (with the blown up region marked by the dotted line) of the  radio-quiet subset  is shown in  the bottom panel.  There is a deficit of radio-quiet objects with $m_{0,agn}<17$, but overall the radio-quiet objects are not systematically fainter.  The lack of very optically bright radio-quiet objects is likely due to a combination of their smaller sample size and their systematically higher redshifts. Thus, we can not attribute the lack of radio-detections for these objects to their having similar radio loudnesses as other \bl\ objects but simply being optically fainter (and thus below the FIRST/NVSS detection limits.)  In addition, even the radio limits placed by their non-detections firmly place them in the radio-quiet regime.  

If we assume the radio-quiet objects are ``normal'' \bl\ objects with comparatively  weak radio emission, then we expect their radio emission to be weak because their SEDs peak in the UV/X-ray (i.e., most should be HBLs, or even extreme HBLs).    We would therefore expect a similar or even larger fraction of radio-quiet objects to have X-ray detections in RASS compared to the radio-loud objects.   However, around 40\% of the radio-loud \bl\ candidates with $m_{0,agn}>17$ have X-ray detections in RASS, and only $\sim$6\% (\noptRquietXray/\noptrq) of the radio-quiet objects have X-ray detections in RASS.  Thus, although the recovered radio-quiet and radio-loud objects share similar optical properties, this hints that  many of the radio-quiet objects might in fact be a distinct population from \bl\ objects (also see \S4.1 of \citealt{shemmer09}, who reach a similar conclusion for high-redshift WLQs).  This, however, does not rule out that some of the radio-quiet objects indeed populate the radio-faint tail of the much larger radio-loud \bl\ distribution.  

We are unaware of any SDSS selection effects that would prevent us from recovering high-redshift radio-loud weak-featured objects or low-redshift radio-quiet weak-featured objects, if such types of objects actually exist.  It is important to note, however, that some of the radio-loud \bl\ candidates lacking spectroscopic redshifts have redshifts bounded by their ``host galaxy'' redshift limits on the lower end, and $z<2.2$ at the higher end.  Thus, it is possible that many of the radio-loud \bl\ candidates actually do have redshifts similar to the bulk of the weak-featured radio-quiet objects.   If that is the case, however, it is then surprising that we find so few radio-loud \bl\ candidates that show even weak-features at $z>1$.   While the redshift distribution of the radio-quiet population is not conclusive evidence on its own that we have actually recovered multiple populations of weak-featured objects, it certainly is suggestive that these might constitute a distinct population from \bl\ objects (especially when taken in context with the unexpectedly small rate of X-ray detections for the radio-quiet objects.)   The lack of a large parent population of radio-loud \bl\ objects at $z>2.2$ does strongly argue against unifying the $z>2.2$ weak-featured radio-quiet objects (i.e., WLQs) with \bl\ objects \citep[e.g., see][]{shemmer09}.

The weak-featured radio-quiet objects may populate a similar region of multiwavelength color space as high-redshift ($z>2.2$) SDSS WLQs \citep[see Figure~5 of][]{shemmer09}.   Among the \noptrq\ weak-featured radio-quiet objects in Tables~\ref{tab:ch4_empdata_rq} and \ref{tab:ch4_deriveddata_rq}, \noptRquietBigZ\ have  $z>2.2$, of which \noptrqWLQ\ have previously been identified as WLQs  \citep{shemmer09,diamond09}.\footnote{Of the other eight $z>2.2$ weak-featured radio-quiet objects, \tabnwlqBL\ are also listed as new WLQ identifications in Table~\ref{tab:ch4_wlq}, another has a tentative redshift $z=2.22$, and the remaining object has a lower limit redshift $z>2.21$ derived from \ion{Mg}{2} absorption.  Neither of the latter two objects show an obvious Ly$\alpha$ forest, so they are not included as new WLQ identifications in Table~\ref{tab:ch4_wlq}. }   
We do not recover all previously known SDSS WLQs because of our stricter $REW$~criterion.  The proposed $z=1.66$ WLQ SDSS J0945+1009 \citep{hryniewicz09_ph} has \ion{Mg}{2} emission with $REW\sim15$~\AA, and even its emission is too large to appear in this sample.  There are likely other similar objects in the SDSS database not included here.

About two-thirds of the radio-quiet weak featured spectra are at $1<z<2.2$.  Have we discovered a large population of lower-redshift analogs to WLQs?     Resolution requires further observations to discriminate from radio-weak \bl\ objects, but either result is interesting.  Polarimetric and variability monitoring, as well as searches for radio and/or X-ray emission would be useful.  Even though the vast majority are not detected by RASS, their extant $\alpha_{ox}$ limits do not yet exclude them from a regime that is typical for \bl\ objects.   It would also be interesting to take ultraviolet spectra of the lower-redshift radio-quiet objects and infrared spectra of the very high redshift radio-quiet objects to compare their emission at similar rest-frames.  

Regardless of their true nature, these radio-quiet objects are a fascinating sub-population:  such a large population of radio-quiet \bl\ objects would pose a serious challenge to the standard unification paradigm that maintains all \bl\ objects are beamed radio galaxies.  Identification of these objects as lower-redshift analogs of high-redshift WLQs would greatly aid our understanding of those other exotic objects.  Just the subset of radio-quiet featureless spectra with $z<2.2$ would be almost as large as the number of $z>2.2$ WLQs previously identified by the SDSS \citep[see][]{diamond09}.  These complementary samples taken together should provide valuable clues to their true nature, as well as insight into the physical conditions necessary to form (or in this case not form) broad line regions near supermassive black holes.   

\section{Summary}
\label{sec:ch4_summary}
We present a sample of \noptall\ objects with weak spectral features selected based on their optical properties.   Approximately 35\% of the objects presented here are new discoveries, and $\sim$75\% were discovered by the SDSS.  Below we summarize our main conclusions:

1.\ Among the tens of thousands of contaminants removed during visual inspection, we note potentially interesting serendipitous discoveries, including 17 unusual white dwarfs,  9 unusual BALQSOs with extreme continuum dropoffs blueward of \ion{Mg}{2}, and \tabnwlq\ high-redshift WLQs (with reliable $z>2.2$ and Ly$\alpha$+\ion{N}{5} $REW<10$~\AA).  Six of these WLQs are additionally classified as \bl\ candidates because their emission features have $REW<5$~\AA.

2.\ Approximately 60\% of our \bl\ candidates have spectroscopic redshifts (or limits from \ion{Mg}{2} absorption).  For objects lacking spectroscopic redshifts, we assume \bl\ host galaxies are standard candles, and we place redshift limits to those objects utilizing the fact that we do not detect host galaxy contamination to their SDSS spectra.  We estimate their ``host galaxy'' redshift limits to be accurate to $\sigma_z = \pm0.064$.  We thus have redshift information (either from weak spectral features or limits placed from the lack of a host galaxy detection) for {\it every} single object in our sample.  We infer the objects lacking spectroscopically-derived redshifts are at systematically higher redshift.  We expect this result, since the objects with the weakest spectral features are arguably among the most highly beamed, which requires a low probability geometry.  So, we are more likely to find the most highly beamed objects at higher redshift, where more volume is probed.  However, there may also be some examples of highly beamed nearby objects in this sample as well.

3.\ Each object emerges with homogeneous multiwavelength data coverage, and we include radio/X-ray fluxes (or limits) for all objects.   Around 80\% of the \bl\ candidates presented here match to a radio source in FIRST or NVSS, and we place upper limits on radio flux densities for the rest (except for \noptallNoRadioInfo\ objects outside the FIRST footprint and not detected by NVSS.)   Approximately 40\% match to an X-ray source in RASS, and we place upper X-ray flux limits for the rest.

4.\ We use each object's high-quality SDSS spectrum to perform spectral decompositions of each object's host galaxy and nucleus, and we report AGN component optical fluxes and luminosities.  

5.\  Based on radio fluxes and limits from the FIRST/NVSS radio surveys, we subdivide our sample into \noptrl\ radio-loud \bl\ candidates and \noptrq\ radio-quiet weak-featured spectra.   There is significant overlap between the optical properties of the radio-loud and radio-quiet objects.  However, the radio-quiet objects tend to be at higher redshift, and they also have a smaller X-ray detection rate.

We argue that  the $\alpha_{ro}<0.2$ objects are a mix of different populations: some of these objects (especially ones lacking spectroscopic redshifts or with small redshifts) probably populate the radio-weak tail of the much larger radio-loud \bl\ population.  Other sources (especially the ones at higher redshifts) are likely WLQs.  The very high-redshift nature of published SDSS WLQs is a selection effect (because they are identified by the presence of the Ly$\alpha$ forest), and  this study provides a systematic search of the SDSS database capable of revealing lower-redshift analogs in large numbers.   

 Even just the $z<2.2$ radio-quiet objects constitute, to our knowledge, the largest sample of such weak-featured radio-quiet objects.  The sample size is larger than many entire venerable \bl\ samples, and this subset is of similar size and complementary to the number of known high-redshift SDSS WLQs.

6.\ The radio-loud objects in general support the standard unification paradigm, which maintains that radio and X-ray selected \bl\ objects populate extreme ends of a single, larger, and continuous \bl\ population.  We recover a mix of LBLs, IBLs, and XBLs, although the SDSS appears biased toward finding HBLs.  However, we note that the large observed scatter in \bl\ properties (e.g., the distribution of their AGN optical spectral indices) is probably due to innate differences as well.  Even just the \noptrl\ radio-loud objects constitute one of the largest samples of \bl\ objects derived from a single selection technique.

7.\   It will be interesting to assemble light curves for these \bl\ candidates once large-scale time-domain surveys, like the  Large Synoptic Survey Telescope \citep[LSST,][]{ivezic08_ph} and  the Panoramic Survey Telescope \& Rapid Response System (Pan-STARRS\footnote{\url{http://pan-starrs.ifa.hawaii.edu}}), come online.   It may even be possible to use flux variability from those surveys as a \bl\ selection criterion.  For example, \citet{bauer09} recently found $\sim$3000 objects in the Palomar-QUEST survey identified by their optical flux variability; many of these objects may be blazars.    Time-domain \bl\ searches over huge areas of the sky  requires telescope access that is unrealistic with standard non-survey telescopes, but well within the capabilities of LSST and Pan-STARRS.

\acknowledgments
We thank the referee for providing excellent suggestions for improving this manuscript.  R.M.P.\ and S.F.A.\ gratefully acknowledge support from NASA/ADP grant
NNG05GC45G. Support for this work was provided by the National Aeronautics
and Space Administration through Chandra Award Number GO9-0126X issued by
the Chandra X-ray Observatory Center, which is operated by the Smithsonian
Astrophysical Observatory for and on behalf of the National Aeronautics
Space Administration under contract NAS8-3060.  W.N.B.\ acknowledges support from NASA LTSA grant NAG5-13035.   A.E.K.\ acknowledges support from NSF grant AST-0507259 to the University of Washington and an NSF Graduate Research Fellowship.  This research has made use of software provided by the Chandra X-ray Center (CXC) in the application package CIAO.  

Funding for the SDSS and SDSS-II
has been provided by the Alfred P.
Sloan Foundation, the Participating Institutions, the National Science
Foundation, the U.S. Department of Energy, the National Aeronautics and
Space Administration, the Japanese Monbukagakusho, the Max Planck Society,
and the Higher Education Funding Council for England. The SDSS Web Site is
\url{http://www.sdss.org/}.  The SDSS is managed by the Astrophysical Research Consortium for the
Participating Institutions. The Participating Institutions are the
American Museum of Natural History, Astrophysical Institute Potsdam,
University of Basel, Cambridge University, Case Western Reserve
University, University of Chicago, Drexel University, Fermilab, the
Institute for Advanced Study, the Japan Participation Group, Johns Hopkins
University, the Joint Institute for Nuclear Astrophysics, the Kavli
Institute for Particle Astrophysics and Cosmology, the Korean Scientist
Group, the Chinese Academy of Sciences (LAMOST), Los Alamos National
Laboratory, the Max-Planck-Institute for Astronomy (MPIA), the
Max-Planck-Institute for Astrophysics (MPA), New Mexico State University,
Ohio State University, University of Pittsburgh, University of Portsmouth,
Princeton University, the United States Naval Observatory, and the
University of Washington.


\end{document}